\documentclass[graybox, nosecnum]{svmult}

\usepackage{natbib}
\usepackage{mathptmx}       
\usepackage{helvet}         
\usepackage{courier}        
\usepackage{type1cm}        
\usepackage{makeidx}         
\usepackage{graphicx}        
\usepackage{multicol}        
\usepackage[bottom]{footmisc}
\usepackage{hyperref}        
\usepackage{soul}            

\makeindex             

\usepackage{bm}
\usepackage{amsmath,amssymb}
\usepackage{mathrsfs}
\usepackage{multirow}

\newcommand{\beq}{\begin{equation}}
\newcommand{\eeq}{\end{equation}}
\newcommand{\bea}{\begin{eqnarray}}
\newcommand{\eea}{\end{eqnarray}}

\newcommand{\nn}{\nonumber\\}
\newcommand{\bs}[1]{\bm{#1}}
\newcommand{\bra}{\langle}
\newcommand{\ket}{\rangle}
\newcommand{\vhat}[1]{\hat{\bm{#1}}}
\renewcommand{\vec}[1]{\bm{#1}}

\renewcommand{\S}[2]{{}^{#1}S_{#2}}

\newcommand{\trip}{\S{3}{1}}
\newcommand{\sing}{\S{1}{0}}

\def\lsim{\mathrel{\rlap{\lower4pt\hbox{\hskip1pt$\sim$}}
    \raise1pt\hbox{$<$}}}         
\def\gsim{\mathrel {\rlap{\lower4pt\hbox{\hskip1pt$\sim$}}
    \raise1pt\hbox{$>$}}}         

\begin{document}
\title*{Theory of Halo Nuclei}
\author{H.-W. Hammer}
\institute{H.-W. Hammer \at Technische Universit\"at Darmstadt, Department of Physics, 64289 Darmstadt, Germany and
ExtreMe Matter Institute EMMI and Helmholtz Forschungsakademie
  Hessen f\"ur FAIR (HFHF), GSI Helmholtzzentrum f\"{u}r Schwerionenforschung
  GmbH, 64291 Darmstadt, Germany}
%
%
\maketitle
\abstract{
  Halo nuclei are characterized by a few weakly bound halo nucleons
and a more tightly bound core. This separation of scales can be
exploited in a few-body description of halo nuclei, since the
detailed structure of the core is not resolved by the halo nucleons.
We present an introduction to the effective (field) theory for low-energy
properties of halo nuclei. The focus is on halos with S-wave interactions
for which universal properties are most pronounced.
The special role of the unitary limit is illustrated using the example of
multineutron systems and the Efimov effect as a universal binding mechanism
for halo nuclei. Connections to ultracold atoms and hadron physics
are highlighted and extensions to higher partial waves, Coulomb forces
and nuclear reactions are briefly touched upon.
}

\section{\textit{Introduction}}
\label{sec:intro}
The emergence of cluster degrees of freedom is an
intriguing aspect of atomic nuclei. 
Certain dripline nuclei form halo states which consist of a tightly bound
core and a few halo nucleons that are only weakly bound to the
core~\citep{Zhukov:1993aw,Hansen:1995pu,Jonson:2004,Riisager:2012it}.
This separation of scales leads to universal properties, which are independent
of the details of the core \citep{Jensen:2004zz,Braaten:2004rn,Hammer:2017tjm}.
These properties are most pronounced in neutron
halos as they are not affected by the long-range Coulomb repulsion between
charged particles.
Neutron halos were discovered in the 1980s at radioactive beam
facilities and are characterized by an unusually large interaction
radius~\citep{Tanihata:2016zgp}, which is directly connected to the
small separation energy of the halo neutrons~\citep{Hansen:1987mc}.

The deuteron is the simplest halo nucleus, consisting of a halo neutron
and a simple proton core. Although there are no emergent cluster degrees of
freedom, it displays the universal features of a halo. In particular,
its wave function extends far beyond the range of
the nuclear force and its root mean square charge radius is about three
times as large as the size of the proton core.
But most halos have a more complex core,
for example the one-neutron halos $^{11}$Be or $^{19}$C~\citep{Jonson:2004}.
Halo nuclei with two valence nucleons 
exhibit three-body dynamics. The case in which the corresponding one-nucleon
halo is beyond the dripline is particularly interesting.
In such \textit{Borromean} three-body systems none of the two-body
subsystems is bound, in analogy to the Borromean rings.

The most carefully studied Borromean two-neutron halo nuclei are
$^6$He and $^{11}$Li \citep{Zhukov:1993aw}.
In the case of $^6$He, the emergent core is a $^4$He nucleus. The two-neutron
separation energy of $^6$He is about 1 MeV, and thus small compared to the
binding and excitation energies of the $^4$He core which are about 
28 and 20 MeV, respectively.

The separation of scales in halo nuclei can formally be exploited using
\textit{effective field theory} (EFT). For a pedagogical introduction
to EFT see, e.g.,~\cite{Kaplan:1995uv,Kaplan:2005es}.
EFT  provides a general framework to calculate the low-energy behavior
of a physical system in an expansion of short-distance over large-distance
scales. The underlying principle is that short-distance physics is not
resolved at low energies and may be included implicitly in 
\textit{low-energy constants}, while long-distance physics must be treated
explicitly. The whole procedure is reminiscent of the multipole expansion
in classical electrodynamics.

Since the relevant energy scales in halo nuclei are so small,
even the pion exchange interaction between nucleons
and/or nuclear clusters is not resolved.
Thus halos can be described by an EFT that contains only
short-range contact interactions, similar to the pionless EFT description
of light nuclei. See,
e.g.,~\cite{Bedaque:2002mn,Epelbaum:2008ga,Hammer:2019poc} for reviews.
For the dynamics of the halo nucleons, the substructure of the core can
also be considered short-distance physics, although low-lying
excited states of the core sometimes have to be included explicitly.
One assumes the core to be structureless and
treats the nucleus as a few-body system of the core and the valence
nucleons.
Corrections from the core structure appear at higher orders in the EFT
expansion, and can be accounted for in perturbation theory.
The philosophy of Halo EFT is similar to that of cluster models
of nuclei~\citep{PhysRev.54.681,PTP-Ikeda,HORIUCHI:1986hpr,Freer_2007}.
But EFT organizes different cluster-model effects into a controlled
expansion based on the scale separation, thereby facilitating the
quantification of theory uncertainties.
A new facet compared to few-nucleon systems is the appearance of
resonant interactions in higher partial waves between the clusters.
This happens, e.g., in the neutron-alpha system which is relevant for $^6$He
and leads to a much richer structure of the EFT
\citep{Bertulani:2002sz,Bedaque:2003wa}.
However, there are many halo nuclei where S-wave interactions are
dominant.

While a field theoretical formalism is advantageous, in particular, when
considering electromagnetic processes and external currents, it is
not strictly necessary. In order to make this introductory
chapter accessible to a wide audience of physicists, we thus use a quantum
mechanical framework, keeping the acronym Halo EFT for convenience.
For an in-depth review of Halo EFT and its applications in the
field theoretical framework as well as a complete bibliography
of previous work, we refer the reader to the review by \cite{Hammer:2017tjm}.

To motivate the Halo EFT approach, we consider a two-body system with
resonant S-wave interactions. The scattering of the core and halo
nucleons at sufficiently low energy is then
determined by their S-wave scattering length
$a$. We consider distinguishable particles of equal mass $m$ and
degenerate pair scattering lengths $a$ for simplicity.
If $a$ is much larger than the range of
the interaction $R$, the system shows universal
properties \citep{Efimov:1971zz,Efimov:1979zz,Braaten:2004rn}.
The simplest example is the existence of a shallow 
two-body bound state or \textit{dimer}
with binding energy and mean square separation
\beq
B_2 = \frac{1}{m a^2}\qquad\mbox{ and }\qquad \langle r^2\rangle =a^2 /2\,,
\label{eq:b2}
\eeq
if $a$ is large and positive. 
(We use natural units with $\hslash=c=1$ throughout this chapter.)
The leading corrections to these universal expressions
are of relative order $R/a$ and can be calculated systematically.
The deuteron binding energy is described by Eq.~\eqref{eq:b2}
to within 35\% accuracy; this improves to 12\% accuracy if the
leading range correction is included.

\begin{figure}[htb]
\centering
\includegraphics[width=6cm,clip=true]{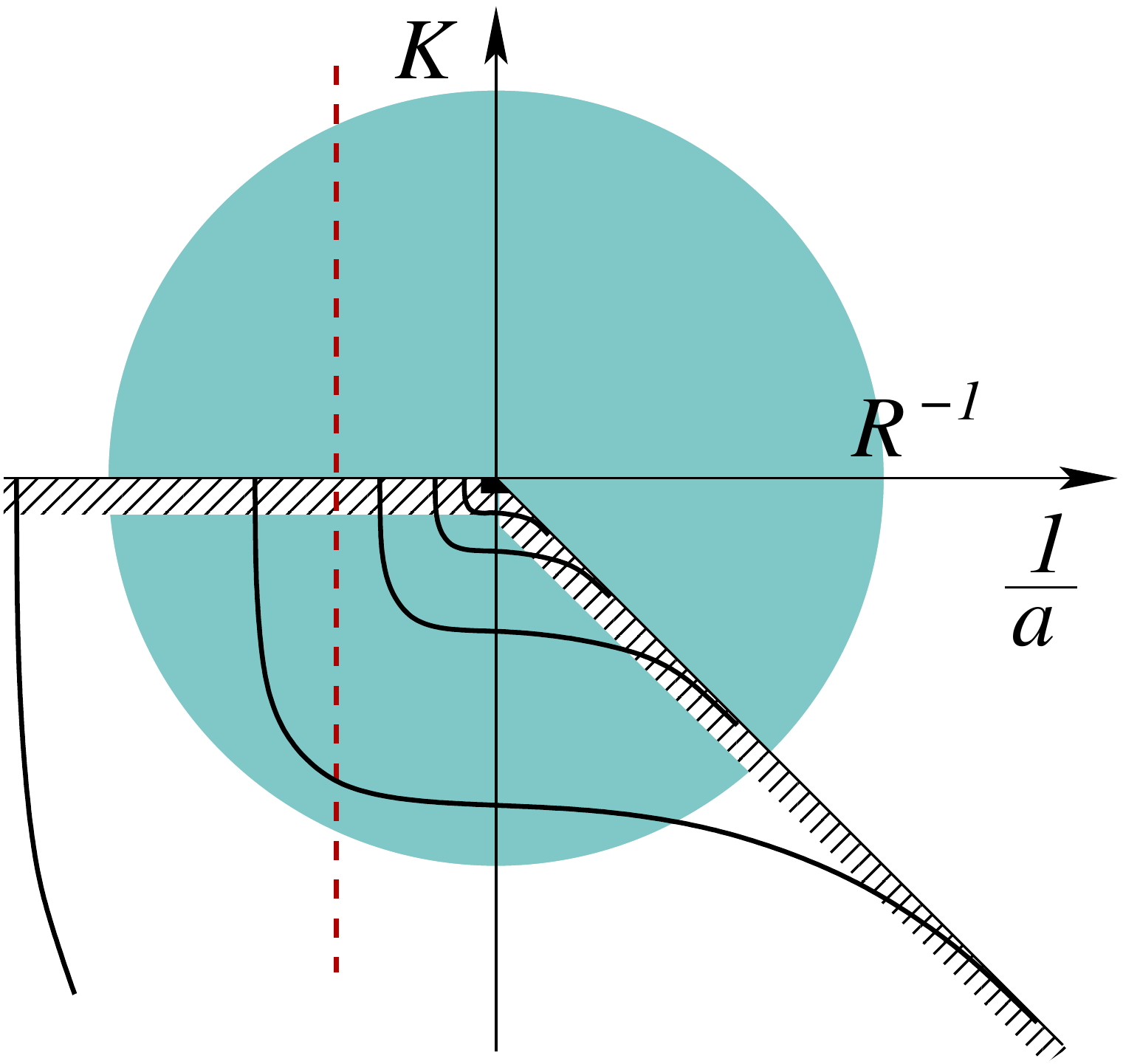}

\caption{Illustration of the Efimov spectrum: The
energy variable $K={\rm sgn}(E)\sqrt{m|E|}$ is shown as a 
function of the inverse scattering length $1/a$.
The shaded circular region exhibits the window of universality.
The solid lines indicate the Efimov states, while the hashed areas
give the scattering thresholds and the dashed vertical line illustrates
an exemplary system with fixed scattering length.
}
\label{fig:efiplot}
\end{figure}

If a third particle is added the two-particle S-wave scattering length no longer
determines the low-energy properties of the system. Observables such as 
the binding energy of three-body bound states and
low-energy scattering phase shifts
are markedly affected by short-distance physics in the three-body
system~\citep{Efimov:1971zz,Bedaque:1998km,Bedaque:1998kg}. 
This additional dynamics can be characterized by a single
\textit{three-body parameter},
$\kappa_*$. All low-energy observables in the
three-body system are functions of $a$ and $\kappa_*$---to leading order
in $R/a$. Moreover, the
Efimov effect \citep{Efimov:1970zz} generates the universal spectrum
of three-body bound states illustrated in Fig.~\ref{fig:efiplot}
in the two-dimensional plane spanned by the momentum variable 
$K={\rm sgn}(E)\sqrt{m|E|}$ and the inverse scattering length $1/a$.
The shaded circular area of radius $R^{-1}$ indicates the
\textit{window of universality} where range corrections are small.
The solid lines indicate the Efimov states while the hashed areas give 
the scattering thresholds below which the bound states can exist. 
The dashed vertical line illustrates an 
exemplary system with a fixed scattering length.
In the \textit{unitary limit} $1/a=0$, the spectrum in Fig.~\ref{fig:efiplot}
reduces to
\beq 
B_3 = e^{-2 \pi n/ s_0}\, \frac{\kappa_*^2}{m}\,,
\label{B3-Efimov-uni}
\eeq
where $s_0=1.00624...$ and the index $n$ labels the three-body states. 

The spectra in Eq.~(\ref{B3-Efimov-uni}) and Fig.~\ref{fig:efiplot}
are invariant under discrete scaling transformations 
by the factor $\lambda_0 = e^{\pi/ s_0}$:
where $n$ is any integer.  This discrete scaling 
symmetry holds for all three-body observables.
It can be seen explicitly in the analytical expression
for the Phillips line at leading order,
\beq
a_{pd}/a = 1.46-2.15 \tan [s_0 \ln(a\kappa_*)+1.06] \,,
\eeq
a correlation between the particle-dimer scattering length $a_{pd}$
and the three-body binding momentum $\kappa_*$. This universal
correlation was originally observed in neutron-deuteron
scattering calculations \citep{Phillips:1968zze}.
The manifestation of discrete scale invariance in observables is
often referred to as \textit{Efimov physics}.

When a fourth particle is added, no new parameters are needed
for renormalization at leading order~\citep{Platter:2004he}.
As a consequence, in the universal regime all four-body
observables are also governed by the 
discrete scaling symmetry
and can be characterized by $a$ and $\kappa_*$.
A similar behavior holds for higher-body observables.
In ultracold atoms, these properties have now been experimentally
verified for up to five particles. 
See \cite{Naidon:2016dpf,Greene:2017cik,Hammer:2019poc} for recent reviews.

The universality of resonant interactions provides the guiding principle for 
the construction of Halo EFT. Thus the description of halos
(and light nuclei, see \cite{Konig:2016utl}) can be organized in an
expansion around the unitary limit.
The breakdown scale $M_{\rm core}$ of this approach is set by the lowest momentum
degree of freedom not explicitly included in the theory. 
The EFT exploits the appearance of a large scattering length
$a \gg 1/M_{\rm core}$, independent of the mechanism generating it. 
In addition to nuclear halo states, examples include ultracold atoms
close to a Feshbach resonance and hadronic molecules in particle physics
\citep{Braaten:2006vd,Hammer:2010kp,Naidon:2016dpf,Guo:2017jvc}.
The typical momentum scale of the theory is $M_{\rm halo} \sim 1/a \sim K$,
which for the systems under consideration here is usually of order tens of MeV.
Meanwhile, the Halo EFT breakdown scale, $M_{\rm \rm core}$,
varies between 50 and 150 MeV, depending on the system.
The expansion is then in powers of $M_{\rm halo}/M_{\rm core}$, and for a
calculation to order $b$ the omitted short-range physics should affect the
EFT's answer by a fractional amount of order
$(M_{\rm halo}/M_{\rm core})^{b+1}$, as long as we consider a process at a
momentum of order $M_{\rm halo}$.
For momenta of the order of the breakdown scale $M_{\rm core}$ the EFT
expansion diverges: the
omitted short-range physics is resolved and has to be treated explicitly.
For many applications, the discussion of uncertainties based on such
estimates is sufficient.
However, a more sophisticated implementation of this
prescription that employs Bayesian statistics to update the size of the
error bar based on the convergence of the perturbative series is 
possible~\citep{Furnstahl:2015rha}.

The direct observation of the discrete scaling symmetry in the
level spectra or reactions of halo nuclei would be a smoking gun for
Efimov physics~\citep{Amorim:1997mq,Jensen:2004zz,Macchiavelli:2015xml},
but the contribution of higher partial waves 
and partial-wave mixing complicate the situation.
While Halo EFT naturally accommodates resonant interactions in 
higher partial waves \citep{Bertulani:2002sz,Bedaque:2003wa},
there is no Efimov effect in this case
\citep{Nishida:2011np,Braaten:2011vf}.
Moreover, universality for resonant P-wave interactions is weaker,
as two parameters, the P-wave scattering volume and effective range,
are required already at leading order in the two-body system.
In higher partial waves this pattern gets
progressively worse~\citep{Bertulani:2002sz,Harada:2007ua}. 
Nevertheless, universality still provides powerful constraints
for the structure and dynamics of halo nuclei~\citep{Hammer:2017tjm}.

The fact that the halo nucleons and the core are treated as
distinguishable particles in Halo EFT means
that the halo nucleons are not antisymmetrized with nucleons in the core---the
latter are not active degrees of freedom in the EFT.
This clearly introduces an error.
However, the contribution of a hypothetical configuration
where a nucleon from the core and from the halo are exchanged
to observables is
governed by the overlap of the wave functions of the core and the halo.
Since the ranges of the core and halo wave functions are
$1/M_{\rm core}$ and $1/M_{\rm halo}$, respectively, the
size of the contribution
is governed by the standard Halo EFT expansion in $M_{\rm halo}/M_{\rm core}$.
Therefore, the impact of anti-symmetrization on observables is controlled
by the Halo EFT expansion
and can be incorporated together with that of other short-distance effects. 
In a single-nucleon halo, these effects enter through the low-energy
constants of the
nucleon-core interaction which is fitted to experimental data or
\textit{ab initio} input.
In a two-nucleon halo, they also enter through a short-range
three-body force. This can be understood as follows: the full
anti-symmetrization of the wave function in a theory with active core
nucleons will result in additional nodes of the halo wave function
since some nucleons must be in excited states to obey the Pauli principle.
In a cluster model, these additional nodes are generated by including
deep unphysical bound states (ghost states) of the core and
the halo, see, e.g.,~\cite{Baye:1985}. In Halo EFT
such deep unphysical states are not included explicitly.
The manifestation of the corresponding physics in Halo EFT can be
understood by assuming that the unphysical states have been integrated
out of the theory. This generates a short-range three-body force between
the core and the two halo nucleons (or modifies an already existing
three-body force in the theory).

Finally, we note that
Halo EFT is not meant to replace \textit{ab initio} approaches to
halo nuclei, instead it complements
\textit{ab initio} approaches by providing universal relations
between different halo observables. Thus it presents a unified
framework for the description of different halo nuclei and their
properties. On the one hand, these universal relations can be 
combined with inputs from \textit{ab initio} theories or experiments to predict
halo properties. On the other hand, they can be used to test 
calculations and/or measurements of different observables for
their consistency.

The chapter is organized as follows. We start by writing down the
effective potential of Halo EFT. This is followed by a discussion of
two- and three-body halo systems with some applications.
Finally, we review universality in multi-neutron
systems and provide pointers to further reading on topics that
have been omitted due to space constraints, including alternative
approaches to halo nuclei.


\section{\textit{Effective potential}}
\label{sec:swavelag}
In order to describe halo nuclei in a non-relativistic EFT framework, it
is important to establish formulae for observables in halos that are
generally suitable for all systems under consideration, and then apply
these expressions to specific cases.  Here, we focus on
one- and two-neutron halos with S-wave interactions between the core and
the valence neutrons.

We introduce an effective potential
$V^{eff}$ to describe a general S-wave halo consisting of
a core ($c$) with spin ${j_c}$ and mass $m_c$ and one or two valence
neutrons ($n$) with spin $1/2$ and mass $m_n$. $V^{eff}$ is written as
a sum of $N$-body potentials, 
\begin{equation}
V^{eff} = V^{eff}_2 + V^{eff}_3 +\ldots\,,
\label{eq:swaveL}
\end{equation}
where the dots stand for higher-body terms not needed here.
The two-body S-wave neutron-neutron ($nn$) and neutron-core ($nc$)
short-range interactions are represented by contact terms.
The two valence neutrons interact in the spin-singlet state, which has a
large negative scattering length.  The neutron and the core couple
into states with total spin ${s}$, whose values can be ${s}_{-}=|{j_c}-1/2|$
and ${s}_{+}={j_c}+1/2$, denoted by $nc(-)$ and $nc(+)$,
respectively. In the case of a spinless core,
the $nc$ interaction forms only one state with ${s}=1/2$.
Due to Galilei invariance, the interaction does not depend on the
center-of-mass momentum. Thus we write
the two-body interaction containing both the $nn$ and $nc$ contact
interactions as
\begin{eqnarray}
  \bra \bs{k}' |V^{eff}_2 |\bs{k}\ket &=&
  \sum_{x=nn,nc(\pm)}\mathscr{P}_{x}\left(
C_{0,x} + C_{2,x} (\bs{k}'^2 +\bs{k})^2/2
+\ldots \right)\,,
\label{eq:Lag-2b}
\end{eqnarray}
where $C_{0,x}$, $C_{2,x}$ are coupling constants and
$\mathscr{P}_{x}$
are projection operators on the corresponding two-body channels
$x=nn,nc(\pm)$.
Moreover, $\bs{k}$ and $\bs{k}'$ are the relative momenta in the
incoming and outgoing channels while the
dots represent higher-order momentum-dependent interaction terms.
A neutron halo can be formed if one of the $nc$ spin channels,
$nc(+)$ or $nc(-)$, has
a scattering length that is much larger than the range of the interaction. 
Then the Efimov scenario from the previous subsection applies.

The three-body effective potential $V^{eff}_3$ does not contribute in a
one-neutron halo nucleus but, in our Halo EFT description of a two-neutron
halo nucleus, arises from the requirement that the three-body problem be
properly renormalized \citep{Bedaque:1998km,Bedaque:1998kg}. 
For simplicity, we write $V^{eff}_3$ using a projection operator
for a specific $n$-$nc$-channel. In an S-wave $2n$ halo system whose
ground state has spin $J$, $V^{eff}_3$ can be represented by a
three-body potential coupling an $nc$ channel and a neutron $n$, 
\begin{equation}
\label{eq:Lag-3b}
\bra\bs{p}'\bs{q}'|V^{eff}_3 |\bs{p},\bs{q}\ket= D_0 \,\mathscr{P}_{n-nc(J)}
+\ldots\,,
\end{equation}
where $\mathscr{P}_{n-nc(J)}$ is the projection operator for the
$n$-$nc$-channel with spin $J$, $D_0$ is the three-body coupling constant,
and $\bs{p}{(')}$, $\bs{q}{(')}$ are Jacobi momenta. 
Based on the effective potential $V^{eff}$, one can calculate low-energy
halo observables. The accuracy of the calculation can be progressively
improved via the systematic expansion in $M_{\rm halo}/M_{\rm core}$ by
adding higher order terms in $V^{eff}$.
In the following we use a shorthand notation, denoting calculations done
at leading, next-to-leading, and order $b$ in this
expansion as LO, NLO, and N$^b$LO.  

\section{\textit{Two-body halos}}
\label{sec:2bamplitude}

The two-body amplitude in a given channel $x=nn,\,nc(\pm)$
is obtained by solving the
Lippmann-Schwinger equation for the effective potential,
\begin{eqnarray}
  \bra \bs{k}'|t_{x}(E)|\bs{k}\ket &=&\bra \bs{k}' |V^{eff}_2 |\bs{k}\ket
  +2\mu_x \int \frac{d^3\bs{q}}{(2\pi)^3} \frac{\bra \bs{k}' |V^{eff}_2 |
    \bs{q}\ket\bra \bs{q}|t_{x}(E)|\bs{k}\ket}{k^2-q^2+i\epsilon}\,,
  \end{eqnarray}
where $E=k^2/(2\mu_x)$ with $k=|\bs{k}|$
is the total kinetic energy in the center-of-mass
frame. The momenta $\bs{k}'$ and $\bs{q}$ are off-shell.
Note that only the part of $V^{eff}_2$ acting in the channel of interest,
$x$, contributes. The reduced masses are $\mu_{nn} =m_n/2$ and
$\mu_{nc(\pm)} =A m_n/(A+1)$, where $A\equiv m_c/m_n$ denotes the
core-neutron mass ratio.  

\begin{figure}[!t]
\centerline{\includegraphics[width=0.8\columnwidth]{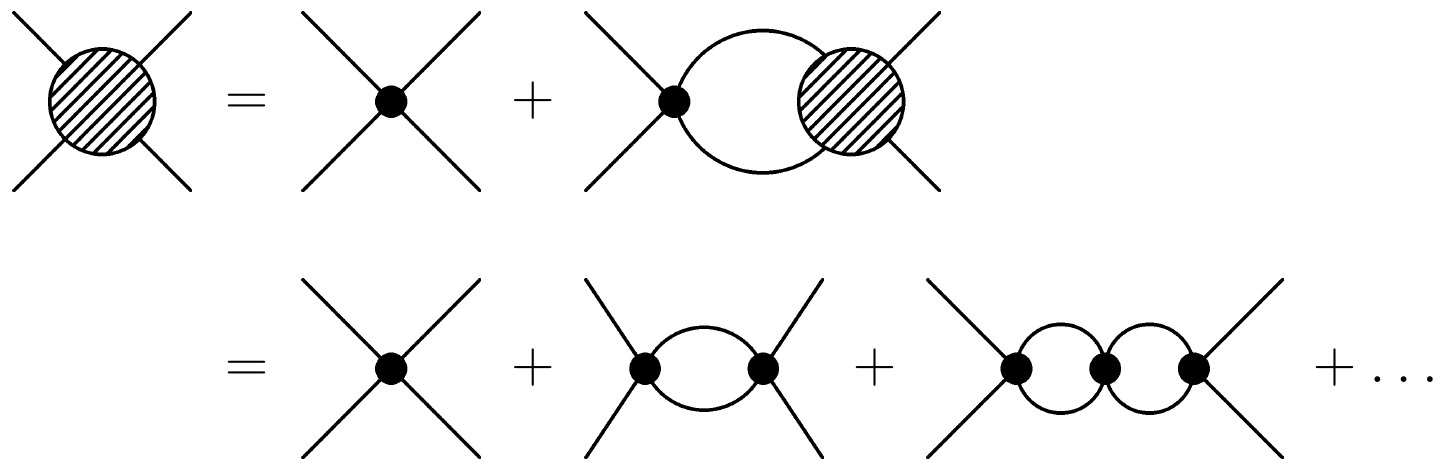}}
\caption{
  Integral equation for the two-body amplitude (1st line) and perturbative
  solution (2nd line).
  The effective potential is represented by the
  filled circle, while the blob indicates the scattering amplitude.}
\label{fig:dsigma-dressed}    
\end{figure}

Inserting the leading order potential we obtain
\begin{eqnarray}
  \label{eq:LS-LO}
  \bra \bs{k}'|t_{x}(E)|\bs{k}\ket &=&C_{0,x}
  +2\mu_x C_{0,x} \int \frac{d^3\bs{q}}{(2\pi)^3} \frac{\bra
    \bs{q}|t_{0,x}(E)|\bs{k}\ket}{k^2-q^2+i\epsilon}\,.
\end{eqnarray}
This integral equation is depicted diagrammatically
in Fig.~\ref{fig:dsigma-dressed}.
Perturbation theory in $C_{0,x}$ leads to a geometric series which
can be summed up to obtain the exact nonperturbative solution:
\begin{eqnarray}
  \bra \bs{k}'|t_{x}(E)|\bs{k}\ket &=&C_{0,x} + C_{0,x} I C_{0,x} +
  C_{0,x} I C_{0,x} I C_{0,x} +
\ldots = [1/C_{0,x} - I]^{-1}\,,
\end{eqnarray}
where
\begin{eqnarray}
  I &=& 2\mu_x \int \frac{d^3\bs{q}}{(2\pi)^3} [k^2-q^2+i\epsilon]^{-1}\,.
\end{eqnarray}
The Integral $I$ is formally divergent and needs to regularized.
This can, e.g., be done by cutting the integral off at high momenta,
using dimensional regularization with power-law divergence subtraction
(PDS)~\citep{Kaplan:1998tg}, or an arbitrary regularization
scheme~\citep{vanKolck:1998bw}.
For simplicity, we apply a sharp momentum cutoff
$\Lambda \gg k$
on the absolute value of $\bs{q}$ and obtain:
\begin{eqnarray}
  I &=& -\frac{\mu_x}{2\pi}\left(ik +\frac{2}{\pi}\Lambda
  + \mathscr{O}(k^2/\Lambda)\right)\,,
\end{eqnarray}
where the terms of $\mathscr{O}(k^2/\Lambda)$ can be dropped.

The resulting amplitude has the same structure as the effective range expansion
(ERE) of the two-body S-wave scattering amplitude in the channel $x$,
\begin{equation}
\bra \bs{k}'|t_{x}(E)|\bs{k}\ket
= \frac{2\pi}{\mu_{x}} \left[\frac{1}{a_{x}} +ik\right]^{-1}\,,
\label{eq:t2b-def}
\end{equation}
with terms of $\mathcal{O}(k^2)$ being omitted. Moreover,
$|\bs{k}|=|\bs{k}'|=k=\sqrt{2\mu_x E}$ in Eq.~\eqref{eq:t2b-def}
is the on-shell relative momentum of the two particles
in the center-of-mass frame.
$a_{x}$ indicates the large S-wave scattering
length in the channel $x$,
which is related to the low-momentum scale by $a_{x} \sim 1/M_{\rm halo}$.
For typical momenta $k \sim M_{\rm halo}$, higher-order corrections
are suppressed by $M_{\rm halo}/M_{\rm core}$.

We use Eq.~\eqref{eq:t2b-def} as a renormalization condition
to determine the running coupling $C_{x}(\Lambda)$ and obtain
 \begin{eqnarray}
   C_{0,x}(\Lambda) &=&  \frac{ 2\pi}{ \mu_{x}} \left[\frac{1}{a_x} -\frac{2}{\pi}
     \Lambda \right]^{-1}\,.
  \label{eq:a-S-wave}
 \end{eqnarray}
 Note also that the unitary term $ik$ does not require
 regularization, since it arises from on-shell intermediate
 states in Eq.~\eqref{eq:LS-LO}.
 
 The momentum-dependent $C_{2,x}$ term from Eq.~(\ref{eq:Lag-2b})
 can be included by writing $C_{0,x} + C_{2,x} (\bs{k}'^2 +\bs{k})^2/2$
 as a two-term separable potential (see, e.g., \cite{Beane:1997pk}
 for details).
 Solving the corresponding Lippman-Schwinger equation,
 we obtain the scattering amplitude
 with effective range term $r_{x}$,
 \begin{equation}
\bra \bs{k}'|t_{x}(E)|\bs{k}\ket
= \frac{2\pi}{\mu_{x}} \left[\frac{1}{a_{x}} -\frac{r_{x}}{2}k^2
  +ik\right]^{-1}\,.
\label{eq:t2b-re}
 \end{equation}
 Note, however, that the Wigner causality bound limits the range
 of cutoffs $\Lambda$ that can be used if one wants to reproduce a positive
 effective range, $r_{x}>0$~\citep{Phillips:1996ae,Hammer:2010fw}.

 For our application to S-wave
 halo nuclei, which exhibit shallow bound or virtual
 states, we are interested in the case of
 a large scattering length, $|a_{x}| \gg |r_{x}|$
 in the channel $x$. The
 scattering amplitude $t_{x}$ can then be expanded around the low-energy
 pole at $k=i\gamma_{x}$, where $\gamma_{x}$ is the binding momentum of
 the two-body S-wave bound state ($\gamma_{x}>0$) or virtual state
 ($\gamma_{x}<0$). At the level of accuracy of Eq.~(\ref{eq:t2b-re}),
 the binding momentum is related to the scattering length $a_{x}$ and the
 effective range $r_{x}$ by 
\begin{equation}
\label{eq:a-gamma}
\frac{1}{a_{x}} = \gamma_{x} -\frac{r_{x}}{2} \gamma_{x}^2\,.
\end{equation}
Therefore, physics of scale $k\sim|\gamma_{x}|$ is enhanced due to the
pole structure of the scattering amplitude. The EFT is constructed
based on a systematic expansion in $\gamma_{x} r_{x}$ or $r_{x}/a_{x}$. 
In the zero-range limit ($r_{x}= 0$) or in the unitary limit
($a_{x}\rightarrow \pm \infty$), we have $\gamma_{x}=1/a_{x}$: the leading order
of the EFT expansion in Eq.~(\ref{eq:a-gamma}) then becomes exact. 

Near the pole, the scattering amplitude, Eq.~(\ref{eq:t2b-re}),
can be expanded about the pole  at $E=-\gamma_{x}^2/(2\mu_{x})$:
\begin{equation}
  \label{eq:D2sigma-pole}
  \bra \bs{k}'|t_{x}(E)|\bs{k}\ket
  = \frac{Z_{x}}{E+\gamma_{x}^2/(2\mu_{x})}+{\rm regular}\,,
\end{equation}
with the residue of the pole
\begin{equation}
  \label{eq:Z-factor}
Z_{x}=
\lim_{E\to -\gamma_{x}^2/(2\mu_{x})} \;\left(E+\frac{\gamma_{x}^2}{2\mu_{x}}\right)
\,t_{x}(E)
= \frac{\pi}{\mu_{x}^2} \frac{2\gamma_{x}}{1-\gamma_{x} r_{x}}\,.
\end{equation}

In a bound two-body system, the residue $Z_{x}$ is connected to the
\textit{asymptotic normalization coefficient} (ANC) $A_x$
of the bound-state wave function.
Near the bound state pole, the full Green's function has the general form
\begin{equation}
\langle \bs{k}' |\frac{1}{E-H} |\bs{k}\rangle
=
\frac{\psi_{x} (\vec{k}^{'})\psi_{x}^*(\vec{k})}{E+\gamma_{x}^2/(2\mu_{x})}
+ {\rm regular}\,,
\label{eqn:ancandT-swave}
\end{equation}
where $\psi_{x}(\vec{k})$ is the asymptotic wave function for the S-wave
bound state in the channel $x$, whose co-ordinate space representation is
\begin{equation}
\label{eq:psi_sigma}
\psi_{x}(\bs{r})=A_{x}  Y_{00}({\hat r}) \frac{\exp(-\gamma_{x} r)}{r}\,,
\end{equation}
with the ANC $A_{x}$ and
$Y_{00}({\hat r})=1/\sqrt{4\pi}$ a spherical harmonic.
To relate $A_{x}$ to $Z_x$, we write the full Green's function
in terms of the free Green's function $1/(E-H_0)$ and $t_{x}(E)$ as
\begin{equation}
\label{eq:G-int}
\frac{1}{E-H}
=\frac{1}{E-H_0}
+ \frac{1}{E-H_0} t_{x}(E) \frac{1}{E-H_0}\,.
\end{equation}
Any bound-state pole can only come from the second piece on
the right-hand side of Eq.~\eqref{eq:G-int}.
Going to the momentum representation and using the expression for
$\bra \bs{k}'|t_{x}(E)|\bs{k}\ket$ near the pole
from Eq.~\eqref{eq:D2sigma-pole} together with
Eqs.~(\ref{eqn:ancandT-swave},~\ref{eq:psi_sigma}), we obtain the relation
\begin{equation}
\label{eq:ANC-swave}
A_{x} = \frac{\mu_x}{\sqrt{\pi}} \sqrt{Z_x} = \sqrt{\frac{2\gamma_{x}}{1-\gamma_{x} r_{x}} }\,.
\end{equation}
Therefore, one can use the ANC $A_{x}$ and the binding momentum
$\gamma_{x}$ to determine the EFT parameters $C_{0,x}$ and $C_{2,x}$,
instead of fixing them from the scattering parameters $a_{x}$ and $r_{x}$.
At LO, $A_{x}=\sqrt{2\gamma_{x}}$ is determined by the binding momentum.
At NLO, the effect of a finite effective range enters and produces an 
ANC ratio different from one: 
\begin{equation}
A_{x}/A_{x,LO}= \left(1-\gamma_{x} r_{x} \right)^{-1/2}\,.
\end{equation}
Note that $A_{x}/A_{x,LO} > 1$ if $\gamma_{x} >0$ (i.e. the
two-body system $x$ is bound) and $r_{x} > 0$. The extent to which this
ratio deviates from one then indicates the importance of range effects.

Although $A_{x}$ is not an observable that is directly measured in
scattering experiments, it can be extracted from such data by
an analytic continuation of the scattering amplitude to negative energies.
There $t_x$ has the pole structure
\begin{equation}
\bra \bs{k}'| t_{x}(E) | \bs{k}\ket =  \frac{2\pi}{\mu_x }  \frac{A^2_{x}/A^2_{x,LO}}{\gamma_{x}+ik} +\rm{regular}\,.
\end{equation}
As compared to the ERE, which is an expansion in powers of
$r_{x}/a_{x}$ around $k=0$, this parameterization in terms of an ANC, dubbed
the \textit{z-parameterization} by~\cite{Phillips:1999hh}, is a more
convenient choice for bound-state calculations. Using this parameterization,
the pole at $k=i\gamma_{x}$ is exactly reproduced at each order, and the
residue of the scattering amplitude, $A^2_{x}/A^2_{x,LO}$, is expanded into a
LO piece $=1$ and an NLO piece $=(A^2_{x}/A^2_{x,LO} -1)$. N$^2$LO and
higher corrections to the ANC are then zero by definition. The
$z$-parameterization of the scattering amplitude  is accurate at
relative order $(M_{\rm halo}/M_{\rm core})^2$, beyond which the shape parameter
enters at $\mathcal{O}(k^4)$.

Here we illustrate the utility of the $z$-parameterization in the calculation
of the matter form factor of one-neutron halos, i.e., we now
choose $x=nc$. The neutron-core 
form factor is the Fourier transform of the coordinate-space probability
density distribution:
\begin{equation}
\label{eq:Fnc-matter}
F_{nc}(|\bs{q}|) 
= \int d^3 r | \psi_{nc}(\bm{r})|^2 \exp(i \bm{q}\cdot \bm{r})\,.
\end{equation}
At LO, we use the zero-range two-body wave function by inserting
$A_{x,LO}=\sqrt{2 \gamma_{x}}$ in Eq.~\eqref{eq:psi_sigma} and obtain 
\begin{equation}
\label{eq:Fnc0-matt}
F_{nc}^{(LO)}(|\bs{q}|) 
= \frac{2\gamma_{nc}}{|\bs{q}|} \arctan\left( \frac{|\bs{q}|}{2\gamma_{nc}}
\right)\,.
\end{equation}

The form factor $F_{nc}^{(NLO)}$ is calculated from
Eq.~\eqref{eq:Fnc-matter} using the full ANC, $A_{nc}$, with an additional
insertion of a constant piece that ensures the matter form factor is properly
normalized~\citep{Phillips:1999hh,Chen:1999tn}, i.e., $F_{nc}(0)=1$.
Consequently, the NLO correction to $F_{nc}$ is
\begin{equation}
\label{eq:Fnc1-matt}
F_{nc}^{(NLO)}(|\bs{q}|) 
= - (A^2_{nc}/A^2_{nc,LO}-1)
\left[1-\frac{2\gamma_{nc}}{|\bs{q}|} \arctan\left( \frac{|\bs{q}|}{2
    \gamma_{nc}}\right)\right]\,.
\end{equation}

Since the low-momentum expansion of the form factor in the one-neutron halo
is related to the mean squared distance between the neutron and the core
$\bra r_{nc}^2 \ket$ via
\begin{equation}
  F_{nc}(|\bs{q}|) = 1-\frac{1}{6} \bra r^2 \ket_{nc}
  \bs{q}^2 + \mathcal{O}(\bs{q}^4)\,,
\end{equation}
we obtain $\bra r^2 \ket_{nc}$ by calculating the first-order derivative
of $F_{nc}$ with respect to $q^2$ at zero. $\bra r^2 \ket_{nc}^{1/2}$ at NLO
is then
\begin{equation}
\label{eq:rnc}
\bra r^2 \ket_{nc}^{1/2} = \frac{A_{nc}/A_{nc,LO}} {\sqrt{2}\gamma_{nc}}\,,
\end{equation}
which reproduces the LO result, Eq.~\eqref{eq:b2}.

With the neutron-core radius in hand we can calculate the matter
radius, which is defined, in the point-nucleon limit, as the average
distance-squared from all nucleons in a halo nucleus to the center of
mass~\citep{Tanihata:2013jwa}:
\begin{equation}
\label{eq:rm-1nhalo}
\bra r_{m}^2 \ket_{1n\rm-halo}  = \frac{A}{(A+1)} \bra r_{m}^2 \ket_{\rm core}
+ \frac{A}{(A+1)^2} \bra r^2 \ket_{nc}\,,
\end{equation}
where the first term is the correction from the matter radius of the core.

A more formal method of keeping the normalization
$F_{nc}(0)=1$ involves imposing gauge invariance of the Lagrangian in the
presence of an external gauge field.
For a discussion of external gauge fields as well as
the form factors of bound states with higher
angular momenta, we refer the reader
to~\cite{Hammer:2011ye,Braun:2018hug}.

\subsection{\textit{Applications 1: bound S-wave neutron halos}}

As an application, we consider some examples of one-neutron S-wave halos,
whose properties are listed in Table~\ref{tab:1n-halo}.
We use $S_{1n}\equiv \textrm{sgn}(\gamma_{nc}) \gamma_{nc}^2/(2\mu_{nc})$ to
denote the neutron-core separation energy in a one-neutron halo,
with $S_{1n}>0$ ($<0$) corresponding to a bound (virtual) S-wave state.

\begin{table}
   \centering
   \begin{tabular}{c c c c c} \hline
           & $^2$H
           & $^{11}$Be   
           & $^{15}$C 
           & $^{19}$C   \\   \hline
           Experiment  &&&&\\  
      $J^P$ 
           & $1^{+}$
           & $1/2^{+}$ 
           & $1/2^{+}$
           & $1/2^{+}$ \\ 
      $S_{1n}$ [MeV] 
           & 2.224573(2)
           & 0.50164(25) 
           & 1.2181(8)
           & 0.58(9) \\ 
      $E^*_{c}$ [MeV] 
           & 293
           & 3.36803(3)
           & 6.0938(2) 
           & 1.62(2) \\      
      $\bra r^2\ket_{nc}^{1/2}$ [fm]
           &  3.936(12)
           &  6.05(23)
           &  4.15(50)
           &  6.6(5)  \\ 
           &  3.95014(156) 
           &  5.7(4)
           &  7.2(4.0)
           &  6.8(7) \\
           &  
           &  5.77(16)
           &  4.5(5)
           &    5.8(3) \\ \hline
      Halo EFT &&&&\\  
      $M_{\rm halo}/M_{\rm core}$
           & 0.33
           & 0.39
           & 0.45
           & 0.6 \\  
      $r_{nc}/a_{nc}$
           & 0.32
           & 0.38
           & 0.43
           & 0.33\\   
      $A_{nc}/A_{nc, LO}$
           &  1.295
           &  1.44
           &  1.63
           &  1.3 \\
      $r_{nc}$ [fm]
           &  1.7436(19)
           &  3.5
           &  2.67
           &  2.6 \\ 
      $\bra r^2\ket_{nc,{\rm theo}}^{1/2}$ [fm]
           &  3.954
           &  6.85
           &  4.93
           &  5.72 \\  \hline
   \end{tabular}
   \caption{Properties of S-wave one-neutron halos.
     One-neutron separation energy $S_{1n}$ taken from
     AME2012~\citep{Audi2012,Wang2012}, while
     lowest core excitation energies $E^*_{c}$ for $A>1$ halos are
     from the TUNL database~\citep{Ajzenberg:1987,Tilley:2004zz,AjzenbergSelove:1991zz}.
     $M_{\rm halo}$ and $M_{\rm core}$ are estimated using $S_{1n}$ and $E^*_{c}$
     except for the deuteron (see text).
     Different experimental values of $\bra r^2\ket_{nc}^{1/2}$
     are given (errors in parentheses). For a discussion of theoretical
     uncertainties, see main text. Compilation taken from
     \citep{Hammer:2017tjm} where also a complete list of references can be
     found.}
   \label{tab:1n-halo}
\end{table}

The deuteron has a spin-triplet ground state ($J^P=1^{+}$), which is
dominated by an S-wave component. The deuteron binding energy as determined
by the 2012 Atomic Mass Evaluation (AME2012)~\citep{Audi2012,Wang2012}
is $S_{1n}=2.224573(2)$ MeV. 
In the language of Halo EFT, the low scale here is $M_{\rm halo}\sim
\gamma_{nc}=\sqrt{m_n S_{1n}}= 45.7$ MeV. An estimate  of the high
scale is obtained from the exchange of pions among nucleons,
$M_{\rm core}\sim M_\pi$.
Well below the pion mass, nuclear potentials can
be considered as short ranged by integrating out the pion degrees of
freedom. Such a \textit{pionless EFT} calculation of the deuteron is then based
on the expansion parameter $M_{\rm halo}/M_{\rm core}\approx 1/3$.
The low energy physics of the deuteron can also be related to $np$
scattering data. In the S-wave spin-triplet channel, the scattering
parameters $a_{\trip}=5.4112(15)$ fm and $r_{\trip}=1.7436(19)$ fm are determined
from an analysis of $np$ elastic scattering data~\citep{Hackenburg:2006qd}.
Their values indicate that the effective range is of expected size
($r_{\trip}\sim 1/M_{\rm core}\sim 1/M_\pi$) while the scattering length is
large, i.e., $a_{\trip}\sim 1/M_{\rm halo}$, leading to an expansion parameter
consistent with the estimate from bound state properties above.

Using Eq.~\eqref{eq:ANC-swave} we obtain the ANC for the deuteron S-wave
wave function to be $A_{\trip} =1.295$~\citep{Phillips:1999hh}. The deuteron
structure radius in the point-nucleon limit, equivalent to
$\bra r^2\ket_{np}^{1/2}/2$, follows from Eq.~\eqref{eq:rnc}
as $1.977$ fm, which overlaps with the value extracted from elastic
electron-deuteron scattering~\citep{Herrmann:1997ia} and agrees with
calculations based on realistic nucleon-nucleon potentials~\citep{Friar:1997js}.

Another example of a one-neutron halo is $^{19}$C, whose ground state was
determined from the Coulomb dissociation spectrum~\citep{Nakamura:1999rp}
to be $J^P=1/2^+$, with a separation energy $S_{1n}=0.53(13)$ MeV between
the $^{18}$C core ($J^P=0^+$) and the last neutron. This result is
consistent with $S_{1n}=0.65(15)$ MeV from one-neutron knock out
reactions~\citep{Maddalena:2001bn}, and $S_{1n}=0.58(9)$ MeV in
AME2012. The first excitation energy of
$^{18}$C is $E^*_{c}=1.62(2)$ MeV~\citep{Ajzenberg:1987}. These values
suggest a separation of low and high scales by $M_{\rm halo}/M_{\rm core}
\sim \sqrt{S_{1n}/E^*_{c}}\approx0.6$. \cite{Acharya:2013nia} performed an
EFT analysis on the $^{19}$C Coulomb dissociation
data~\citep{Nakamura:1999rp,Nakamura:2003cyk}.
They extracted $A_{nc}/A_{nc,LO}=1.31$, together with
$S_{1n}=0.575(55)(20)$ MeV---the latter in agreement with the extraction
by~\cite{Nakamura:1999rp} and AME2012.
These correspond to ERE parameters $a_{nc}=7.75(35)(30)$ fm and
$r_{nc}=2.6^{+0.6}_{-0.9}\pm0.1$ fm, where the first error indicates
the statistical uncertainty from the fit to data and the second
one quantifies the systematic N$^3$LO EFT uncertainties, which are
estimated to be of relative size $(r_{nc}/a_{nc})^3$.
In fact, the ratio $r_{nc}/a_{nc}=0.33$ suggests that
the EFT may converge faster than the naive dimensional estimate,
$M_{\rm halo}/M_{\rm core} \approx 0.6$.

The above results for ${}^{19}$C imply an S-wave binding momentum
$\gamma_{nc}=32.0$ MeV. This, together with the extracted ANC,
yields the neutron-core distance from Eq.~\eqref{eq:rnc} to be
$\bra r^2\ket_{nc}^{1/2}=5.72$ fm, which agrees with values deduced by the
E1 sum rule of Coulomb dissociation~\citep{Nakamura:1999rp} and extracted
from the charge-changing cross section~\citep{Kanungo:2016tmz} measurements
(see Table \ref{tab:1n-halo}). 

Other examples of one-neutron halos are $^{11}$Be and $^{15}$C. Their ground
states both have spin-parity quantum numbers $J^P=1/2^{+}$, with one valence
neutron attached to the $^{10}$Be and $^{14}$C cores ($J^P=0^{+}$). The
one-neutron separation energies $S_{1n}$ of $^{11}$Be and $^{15}$C are
given in the atomic mass evaluation AME2012; while
the first excitation energies of the cores $E_c^*$ are obtained from the
TUNL data base~\citep{Tilley:2004zz,AjzenbergSelove:1991zz},
cf.~Table \ref{tab:1n-halo}. 

Based on naive dimensional analysis the EFT expansion parameter in $^{11}$Be
is $M_{\rm halo}/M_{\rm core} \sim \sqrt{S_{1n}/E^*_{c}}=0.39$. The ANC in
the $^{11}$Be ground state was obtained in an \textit{ab initio}
calculation that used the No-Core Shell Model with Continuum approach
as $A_{nc}/A_{nc,LO}=1.44$~\citep{Calci:2016dfb}. This corresponds to
$r_{nc}=3.5$ fm, which yields $r_{nc}/a_{nc}=0.38$ in $^{11}$Be, in agreement
with the expansion parameter inferred from  $\sqrt{S_{1n}/E^*_{c}}$.

EFT calculations for $^{15}$C were performed in
\citep{Rupak:2012cr,Fernando:2015jyd}. By fitting the Halo EFT neutron
capture cross section to experiment~\citep{Nakamura:2009zzc}, they
determined the ANC ratio $A_{nc}/A_{nc,LO}=1.63$, or equivalently,
an effective range $r_{nc}=2.67$ fm. Their calculation suggested an
unnaturally scale for the effective range, with
$r_{nc}\sim 1/M_{\rm halo}$, implying that the
z-parameterization, $A_{nc}^2/A_{nc,LO}^2-1$, becomes non-perturbative
in this system.
Using the ANC extracted in \citep{Rupak:2012cr}, we obtain
$r_{nc}/a_{nc}=0.43$, which, despite the somewhat large effective range,
is still consistent with $M_{\rm halo}/M_{\rm core}\approx 0.45$.

\subsection{\textit{Applications 2: unbound S-wave neutron halos}}
Halo-like features also exist in unbound systems, if such systems display
a large negative scattering length. In the $np$ spin singlet state, the
ERE parameters $a_{\sing(np)}= -23.7148(43)$ fm and $r_{\sing(np)} =2.750(18)$
fm are determined from at analysis of low-energy $np$ elastic-scattering
data~\citep{Hackenburg:2006qd}. The $nn$ singlet state is also
unbound, with a scattering length $a_{\sing(nn)}= -18.6(5)$ fm 
and $r_{\sing(nn)} =2.83(11)$ fm 
obtained from the neutron time-of-flight spectrum in
radiative pion capture on the deuteron~\citep{Chen:2008zzj}.
It should be noted, however, that there is a systematic and significant
difference between the extracted values of $a_{\sing(nn)}$ from
neutron-induced deuteron breakup reactions measured by two different
collaborations with different experimental
setups~\citep{Gardestig:2009ya}. A Halo EFT
analysis of the reaction $^{6}$He$(p,p\alpha)nn$ in inverse kinematics
at high energies may help to resolve this discrepancy~\citep{Gobel:2021pvw}.
Further below we will also discuss some universal
features of unbound multineutron systems based on the fact that
$r_{\sing}/a_{\sing}$ is a small parameter \citep{Hammer:2021zxb}.

$^{11}$Li, whose ground state has spin-parity $J^P=3/2^-$ was one of the
first halo nuclei~\citep{Tanihata:1986kh} beyond the few-nucleon systems to
be discovered. $^{11}$Li is a Borromean two-neutron halo, where the
neutron-core is unbound. Here we focus on the separation of scales in
$^{10}$Li and refer to later sections for properties of $^{11}$Li. The
ground state of the $^9$Li core has $J^P=3/2^-$ and a first excitation energy
$E^*_{c}=2.691(5)$ MeV~\citep{Tilley:2004zz}, which sets $M_{\rm core}$.
The scale $M_{\rm halo}$ is associated with the $^{10}$Li ground state, which
can be interpreted as
an unbound S-wave neutron-core virtual state ($J^P=1^-$ or $2^-$) with
$S_{1n}=-25(15)$ keV~\citep{Tilley:2004zz}. The EFT expansion parameter for
$^{10}$Li is estimated as $M_{\rm halo}/M_{\rm core}\sim \sqrt{|S_{1n}|/E^*_{c}}
\approx 0.1$. A proton removal reaction experiment~\citep{Smith:2015bpa}
observed two resonance states of $^{10}$Li at energies $E_{2,1+}=110(40)$ keV
and $E_{2,2+}=500(100)$ keV above the neutron-core threshold. These are
expected to be P-wave states. As such they enter at higher orders in the EFT
compared to the S-wave virtual state, whose large scattering
length promotes it to LO. 

$^{21}$C is another unbound neutron-core system~\citep{Langevin:1985ior}.
The ratio between the one-neutron separation energies of
$^{21}$C and $^{20}$C from AME2012 provides a valid expansion parameter
$\lsim 0.18$.
The neighboring isotope $^{22}$C has recently been identified as a
weakly-bound two-neutron halo and is the dripline nucleus of carbon
isotopes. A Glauber-model analysis of the reaction cross section of
$^{22}$C on a hydrogen target~\citep{Tanaka:2010zza} and a measurement of
the two-neutron removal reaction on $^{22}$C~\citep{Kobayashi:2011mm}
suggest that $n-^{20}$C is preferentially in $1/2^{+}$ configuration.
The two-neutron halo structure of $^{22}$C implies that $^{21}$C occupies
an S-wave virtual state near the unitary limit. However, a recent study of
the $n-^{21}$C decay spectrum via one-proton removal from the
$^{22}$N beam~\citep{Mosby:2013bix} implies that the $n-^{20}$C scattering
length is not large, $|a_{nc}|<2.8$ fm (or equivalently $S_{1n}<-2.8$ MeV).
Therefore, further studies on the properties of $^{21}$C are needed.

\section{\textit{Three-body halos}}
\label{sec:3Bswave}

Here we consider two-neutron halos as a neutron-neutron-core three-body
system. We use the Jacobi momentum plane-wave state
$|\boldsymbol{p},\boldsymbol{q}\ket_i$ to represent the kinematics of the
three-body system in the center-of-mass frame. The index $i$ indicates
that these momenta are defined in the two-body fragmentation channel $(i,jk)$,
in which particle $i$ is the spectator and  $(jk)$ the interacting pair. 
Based on this definition, $\boldsymbol{p}$ represents the relative momentum
in the pair $(jk)$; while $\boldsymbol{q}$ denotes the relative momentum
between the spectator $i$ and the $(jk)$ pair.
The plane-wave states are normalized as~\citep{Glockle:1983}:
\begin{equation}
{}_i\bra \bs{p} \bs{q} |
\bs{p}' \bs{q}'\ket_i = (2\pi)^6 \delta^{(3)}(\bs{p}-\bs{p}')
\delta^{(3)}(\bs{q}-\bs{q}')\,.
\label{eq:Jacobi-1}
\end{equation}
The Jacobi momenta are related to the momenta in the direct product of
three single-particle states $|\bs{k}_1,\bs{k}_2,\bs{k}_3\ket$
in the center-of-mass frame (i.e., $\bs{k}_1+\bs{k}_2+\bs{k}_3=0$) by
\begin{eqnarray}
\bra \bs{k}_1,\bs{k}_2,\bs{k}_3 |  \bs{p} \bs{q}\ket_i &=& 
(2\pi)^6 \delta^{(3)}\left(\bs{p}_i -\mu_{jk}\left[\frac{\bs{k}_j}{m_j}
  -\frac{\bs{k}_k}{m_k}\right]\right)\,\nonumber \\
&&\qquad \times
\delta^{(3)}\left(\bs{q}_i -\mu_{i(jk)}\left[\frac{\bs{k}_i}{m_i}
  -\frac{\bs{k}_j+\bs{k}_k}{M_{jk}}\right]\right)\,,
\label{eq:Jacobi-2}
\end{eqnarray}
where $M_{jk}=m_j+m_k$, $\mu_{jk}=m_j m_k /M_{jk}$, and
$\mu_{i(jk)}= m_i M_{jk}/(m_i+M_{jk})$ are mass parameters.

To discuss the spin and parity of a halo nucleus, we introduce the
partial-wave-decomposed representation. The relative orbital angular
momentum and the spin of the pair $(jk)$ are defined as $l_i$ and $s_i$.
They are coupled to form the total angular momentum $j_i$ in the pair.
We also define the spin of the spectator $i$ as $\sigma_i$, the relative
orbital angular momentum between the spectator $i$ and the pair $(jk)$ as
$\lambda_i$, and the corresponding total angular momentum as $I_i$. The
overall orbital angular momentum, spin and total angular momentum of the
three-body system are denoted by $L_i$, $S_i$ and $J$. We then have:
\begin{eqnarray}
\bs{L}_i =& \bs{l}_i+\bs{\lambda}_i\,, \quad
\bs{S}_i = \bs{s}_i+\bs{\sigma}_i\,, \quad
\label{eq:J-LS}
\bs{J} = \bs{L}_i+\bs{S}_i = \bs{j}_i+\bs{I}_i\,.
\end{eqnarray}

Knowing the spin and orbital-angular-momentum quantum numbers, we can
construct three-body eigenstates with respect to the spin and
orbital-angular-momentum operators. Note that $J$ is a conserved quantum
number, which is independent of the choice of partition representations given
in Eq.~\eqref{eq:J-LS}. We decompose the Jacobi momenta 
with respect to these spin and orbital- and total-angular-momentum quantum
numbers by~\citep{Glockle:1983}
\begin{equation}
\label{eq:quantum-number}
\left|p,q;\Omega_i\right\rangle_{i} = \sum\limits_{L_i S_i}\sqrt{\widehat{j}_i \widehat{I}_i \widehat{L}_i \widehat{S}_i}\,
\left\{\begin{array}{ccc}
  l_i & s_i &  j_i \\
  \lambda_i & \sigma_i &  I_i \\
  L_i     & S_i     & J 
\end{array}\right\}\,
\left| p,q;(l_i,\lambda_i)L_i\,; \left(s_i,\sigma_i\right)S_i\,;(L_i S_i)J \right\rangle_{i}\,,
\end{equation}
where $\widehat{j}_i$ denotes $2{j}_i+1$ (the same holds for
$\widehat{I}_i$, $\widehat{L}_i$ and $\widehat{S}_i$), $p\equiv|\bs{p}|$,
and $q\equiv|\bs{q}|$. The collective symbol $\Omega_i$ represents all
conserved spin, orbital- and total-angular-momentum quantum numbers in the
partition  $(i, jk)$.

In this chapter, we focus on S-wave two-neutron halos,
where $l_i=\lambda_i=L_i=0$ and the values of the total angular momenta are
equal to their corresponding spins. Therefore, one can use the spins alone
to represent the decomposed plane-wave state in S-wave $2n$ halos as 
$| p,q; (s_i,\sigma_i)S\rangle_{i}$, where the three-body total spin $S$ is the
same in different partitions. In the $(c,nn)$ partition, $S=\sigma_c$ since the
two-neutron pair is spin singlet ($s_c=0$). Therefore, in the $(n,nc)$
partition, the neutron-core pair with spin $s_n=|\sigma_c\pm1/2|$ couples with
the second neutron with spin $\sigma_n=1/2$ to form the three-body total spin
$S=\sigma_c$.

\begin{figure}[!t]
\centerline{\includegraphics[width=0.8\columnwidth]{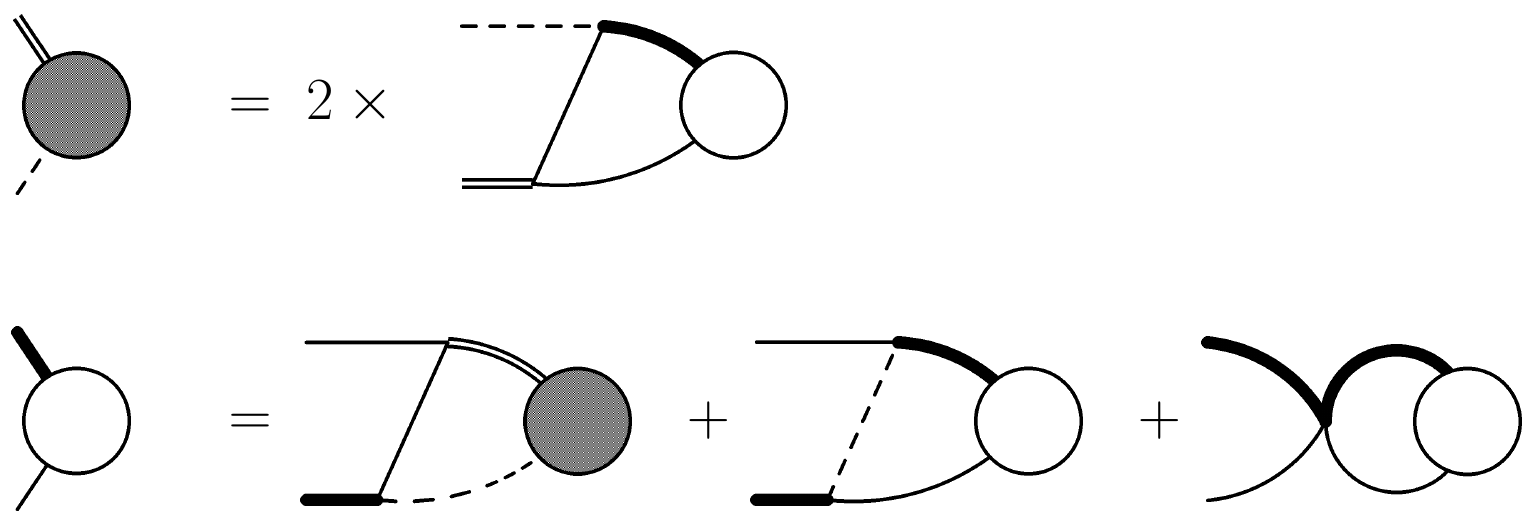}}
\caption{
  The integral equations for the transition amplitudes $\mathcal{A}_c$
  (shaded circle)  and $\mathcal{A}_n$ (empty circle).
   The solid (dashed) lines represent neutrons (core),
  while the double (thick solid) lines represent $nn$ ($nc$) scattering
  amplitudes, respectively. The last piece
  represents the contribution from the $nnc$ three-body force.}
\label{fig:faddeev-eq}       
\end{figure}

Moreover, we assume that the neutron-core states with $s_n=|\sigma_c\pm1/2|$
are degenerate and have equal scattering lengths. Under this assumption,
the three-body formalism in S-wave halos with a spin-zero core becomes
general for an arbitrary S-wave two-neutron halo with spin $\sigma_c$. 
We use the Faddeev formalism~\citep{Faddeev:1960su,Glockle:1983,Afnan:1977pi}
and decompose the three-body wave function into components.
In Halo EFT, $2n$-halo nuclei are described by the
transition amplitudes, $\mathcal{A}_c$ and $\mathcal{A}_n$, connecting the
spectator and the interacting pair to the three-body bound state.
$\mathcal{A}_c$ and $\mathcal{A}_n$ are represented by functions of the
Jacobi momentum $\bs{q}$ between the spectator and the pair, and are the
solution of coupled-channel homogeneous integral
equations~\citep{Canham:2008jd,Acharya:2013aea} which are illustrated
with Feynman diagrams in Fig.~\ref{fig:faddeev-eq}. 

At leading order, one can take $r_{x}\rightarrow 0$ for $x=nn, nc$ and
obtains \citep{Hammer:2017tjm},
\begin{eqnarray}
\mathcal{A}_c(\bm{q}) &=& 2 \int \frac{d^3 q}{4\pi^2} \, G_0^n
\left(\pi_2(\bm{q}',\bm{q}),q;B_3\right) \,
\tau_{nc}(q';B_3) \, \mathcal{A}_n(\bm{q}')\,,
\nonumber\\
\label{eq:inteq-An}
\mathcal{A}_n(\bm{q}) &=& \int \frac{d^3 q'}{4\pi^2} \, G_0^n
\left(\pi_2(\bm{q},\bm{q}'),q';B_3\right) \,
\tau_{nn}(q';B_3) \, \mathcal{A}_c(\bm{q}')
\nn
&&+\int \frac{d^3 q'}{4\pi^2} \, \left[G_0^c \left(\pi_3(\bm{q}',\bm{q}),q;
  B_3\right) +\frac{H(\Lambda)}{\Lambda^2} \right]\,
\tau_{nc}(q';B_3) \, \mathcal{A}_n(\bm{q}')\,,
\end{eqnarray}
where $G_0^c$ and $G_0^n$ are the three-body Green's functions expressed in two different partitions:
\begin{eqnarray}
G_0^c(p,q;B_3) &=& \left( m_n B_3 + \frac{A+1}{2A} p^2 + \frac{A+2}{2(A+1)} q^2 \right)^{-1}\,,\nonumber
\\
G_0^n(p,q;B_3) &=& \left( m_n B_3 +  p^2 + \frac{A+2}{4A} q^2 \right)^{-1}\,.
\end{eqnarray}
The momentum variables
$\bs{\pi}_1$, $\bs{\pi}_2$ and $\bs{\pi}_3$ are defined as
\begin{eqnarray}
\bm{\pi}_1(\bm{q},\bm{q}') = \bm{q} + A \bm{q}'/(A+1)\,,
\nonumber\\
\bm{\pi}_2(\bm{q},\bm{q}') = \bm{q} + \bm{q}'/2\,,
\nonumber\\
\bm{\pi}_3(\bm{q},\bm{q}') = \bm{q} + \bm{q}'/(A+1)\,.
\end{eqnarray}
Moreover, the functions $\tau_x(q;B_3)$ with $x=nn,nc$ in the three-body integral
equations~\eqref{eq:inteq-An} are rescaled two-body scattering amplitudes
from Eq.~\eqref{eq:t2b-def},
\begin{eqnarray}
\tau_{nn}(q;B_3) 
&=& \frac{2}{-\gamma_{nn} + \sqrt{m_n B_3 + \frac{A+2}{4A} q^2 }}\,,
\nn
\tau_{nc}(q;B_3) 
&=&   \frac{(A+1)/A}{-\gamma_{nc} + \sqrt{ \frac{A}{A+1}
    \left(2 m_n B_3 + \frac{A+2}{A+1} q^2 \right)}}\,,
\end{eqnarray}
while $H(\Lambda)$ is a dimensionless three-body force parameter
that emerges from including the three-body interaction
$D_0 \propto m_n C_{0,nc}^2  H(\Lambda)/\Lambda^2$ from
Eq.~\eqref{eq:Lag-3b}. Note that our amplitudes $\mathcal{A}_c$ and
$\mathcal{A}_n$ correspond to the amplitudes $\tilde{\mathcal{A}}_c$ and 
$\tilde{\mathcal{A}}_n$ in \citep{Hammer:2017tjm}.

By projecting the three-body transition amplitudes to S-waves, we
simplify the integral equations to the expressions given
in~\citep{Canham:2008jd,Acharya:2013aea}
\begin{eqnarray}
  \mathcal{A}_c(q) &=& \frac{2}{\pi} \int^\Lambda_0 d q' \, q'^2 K^n(q',q;B_3)
  \, \tau_{nc}(q';B_3) \, \mathcal{A}_n(q')\,, 
\nonumber \\
\mathcal{A}_n(q) &=& \frac{1}{\pi}  \int^\Lambda_0 d q' \, q'^2 K^n(q,q';B_3)
\, \tau_{nn}(q';B_3) \, \mathcal{A}_c(q')
\nn
&&+\frac{1}{\pi}  \int^\Lambda_0 d q' \, q'^2 \left[K^c(q',q;B_3)
  +\frac{H(\Lambda)}{\Lambda^2} \right]\,
\tau_{nc}(q';B_3) \, \mathcal{A}_n(q')\,,
\label{eq:faddeevswave2}
\end{eqnarray}
where a sharp momentum cutoff $\Lambda$ has been applied
as a regulator. This ensures that the integral equations
\eqref{eq:faddeevswave2} have a unique solution.
The kernel functions are
\begin{eqnarray}
K^n(q,q';B_3) 
&=& \frac{1}{2} \int_{-1}^{1} d (\vhat{q}\cdot\vhat{q}') G_0^n
\left(\pi_2(\bm{q},\bm{q}'),q';B_3\right)
= - \frac{1}{qq'} \textrm{Q}_0 (z_{nc})\,, 
\nonumber
\\
K^c(q,q';B_3) 
&=& \frac{1}{2} \int_{-1}^{1} d (\vhat{q}\cdot\vhat{q}')
G_0^c \left(\pi_3(\bm{q},\bm{q}'),q';B_3\right)
= - \frac{A}{qq'} \textrm{Q}_0 (z_{nn})\,,
\label{eq:Xnn-s}
\end{eqnarray}
with $\textrm{Q}_l$ the Legendre function of the second kind, which is
related to the Legendre polynomial $P_l$ by $\textrm{Q}_l(z) \equiv
\frac{1}{2} \int^1_{-1} dx\, P_l(x)/(z-x)$.
The arguments $z_{nn}$ and $z_{nc}$ of $\textrm{Q}_0$ in
Eqs.~\eqref{eq:Xnn-s} are defined as
\begin{eqnarray}
\label{eq:znn-znc}
z_{nc} &=& -\frac{1}{qq'} \left(m_n B_3 + q^2 + \frac{A+1}{2A} q'^2\right)\,,\nn
z_{nn} &=& -\frac{A}{qq'} \left(m_n B_3 + \frac{A+1}{2A} (q^2+q'^2)\right)\,.
\end{eqnarray}
For bound states $B_3>0$, so $z_{nc},z_{nn}< -1$ and no singularity
of $\textrm{Q}_0$ is encountered. The superscript $y=n,c$ in the kernel
functions $K^y$ indicates the particle exchanged in the one-particle
exchange interaction, see Fig.~\ref{fig:faddeev-eq}.

\begin{figure}[!t]
\centerline{\includegraphics[width=0.75\columnwidth]{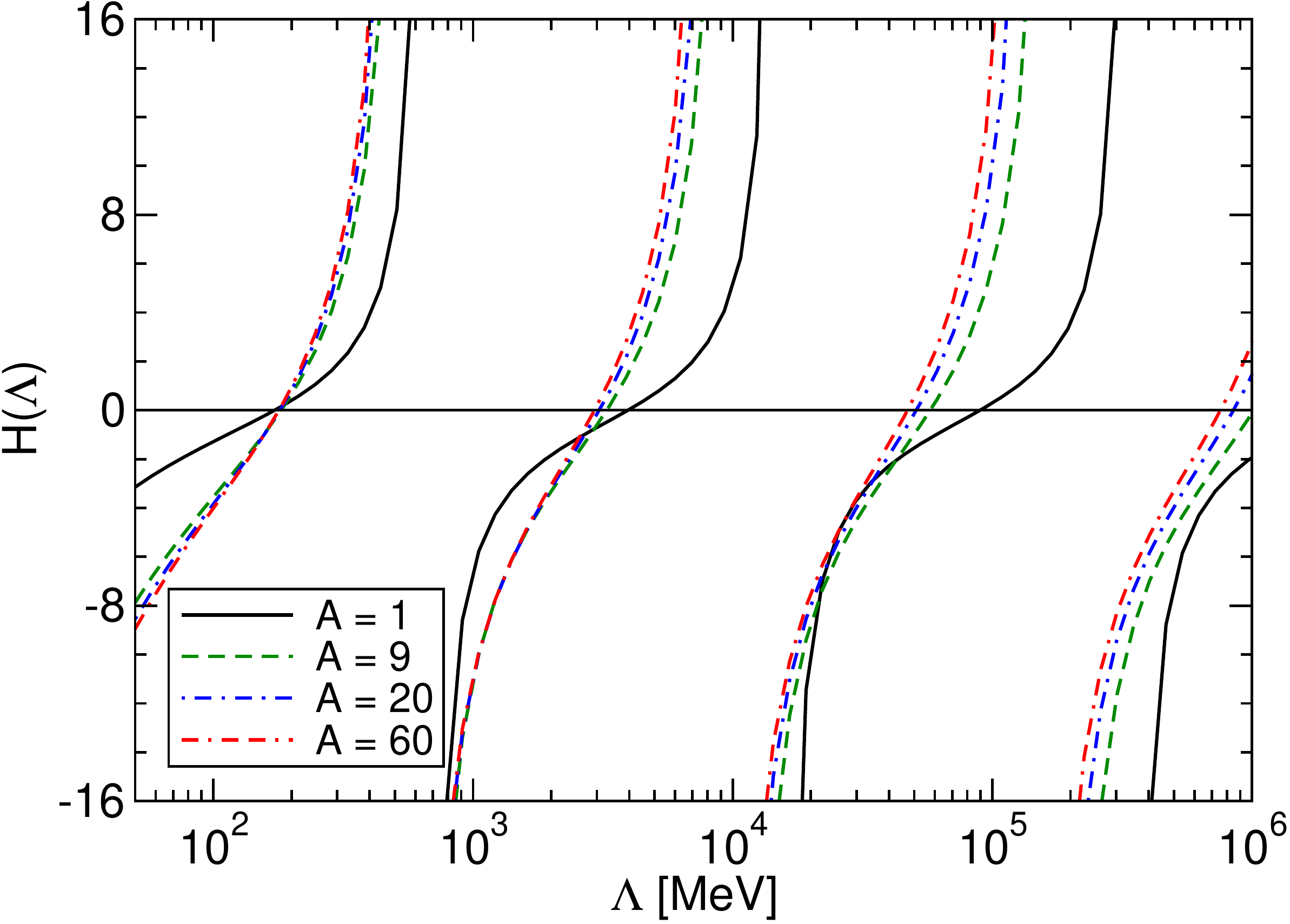}}
\caption{
  (Color online) The running of $H(\Lambda)$ as a function of the cutoff
  $\Lambda$ for systems with $A=1$ (solid), $A=9$ (dashed),
  $A=20$ (dash-dotted), and $A=60$ (dash-dash-dotted). Numerical results
  are from \citep{Hammer:2017tjm} and obtained for $\gamma_{nn},\gamma_{nc}=0$
  and $B_3=1$ MeV.}
\label{fig:H03b-A}  
\end{figure}

To solve the coupled integral equations, we look for an energy $E=-B_3$
where the eigenvalue of the integral-equation kernel is one.  To keep
low-energy observables invariant
under changes of the regulator $\Lambda$ the parameter $H(\Lambda)$
is tuned so that one three-body observable, such as the binding energy $B_3$,
is kept fixed as $\Lambda$ is varied. In two-neutron halos with three pairs
resonantly interacting in S-waves, the resulting asymptotic running of
$H$ is characterized by a limit cycle~\citep{Wilson:1970ag,Albeverio:1981zi,Bedaque:1998km,Bedaque:1998kg,Barford:2004fz,Mohr:2005pv}.
In particular, the discrete scale invariance of this problem in the
ultra-violet results in $H(\Lambda)$ being a log-periodic function of $\Lambda$.The numerical results for various $A$ are well described by
\citep{Hammer:2017tjm}
\begin{equation}
\label{eq:H0-3bA}
H(\Lambda) = c_A \frac{ \sin\left(s_0\log(\Lambda/\Lambda_*) + \arctan(s_0) +d_A \right) }
{ \sin\left(s_0\log(\Lambda/\Lambda_*) - \arctan(s_0) \right) } + e_A\,.
\end{equation}
Here $c_A$, $d_A$, and $e_A$ are constants that depend on the
core/neutron mass ratio $A$, with $d_A=e_A=0$ when $A=1$. The 
renormalization parameter $\Lambda_*$ is determined by an observable in a
given two-neutron halo. It is proportional to the binding 
momentum in the unitary limit, $\kappa_*$, modulo factors of
the limit cycle period $\exp(\pi/s_0)$. 
Here, $s_0$ is a solution of a transcendental
equation~\citep{Nielsen2001,Braaten:2004rn}:
\begin{equation}
\label{eq:s0-3bA}
\cosh^2\left(\frac{\pi s_0}{2}\right)
-\cosh \left(\frac{\pi s_0}{2}\right)
 \frac{2\sinh(\theta_{1}s_0)}{s_0\sin(2 \theta_{1})}
-\frac{8 \sinh^2(\theta_{2}s_0)}{s_0^2\sin^2(2 \theta_{2})} = 0\,,
\end{equation}
where $\theta_{1}=\arcsin(1/(1+A))$, $\theta_{2}=\arcsin\sqrt{A/(2+2A)}$.
Eq.~\eqref{eq:H0-3bA} was first derived in systems of three equal-mass
particles~\citep{Bedaque:1998km,Bedaque:1998kg}, where $s_0=1.00624$ is
obtained from Eq.~\eqref{eq:s0-3bA}. This corresponds to a discrete scaling
factor $\exp(\pi/s_0)=22.694$ and reveals the presence of the Efimov effect
\citep{Efimov:1971zz}. 
The log-periodicity of $H(\Lambda)$ persists when the core and neutron masses
are not equal. 
The running of $H(\Lambda)$, shown in Fig.~\ref{fig:H03b-A}, clearly
indicates that the limit-cycle behavior is present for values of $A > 1$.
The numerical results shown there are from \citep{Hammer:2017tjm} and
were obtained for the unitary limit, $\gamma_{nn}=\gamma_{nc}=0$, and
$B_3=1$ MeV. In general, the log periodicity persists also for
finite binding momenta if $\Lambda \gg \gamma_{nn},\gamma_{nc}$.

The full three-body wave function $\Psi$ can 
then be obtained by connecting  the three-body transition amplitudes
with external one-body propagators and two-body scattering amplitues.
The wave function can be represented in two different Jacobi partitions
labeled by the spectator $n$ or $c$. In  S-wave two-neutron halos, we
obtain~\citep{Canham:2008jd,Acharya:2013aea}
\begin{eqnarray}
\Psi_n(p,q)
&=&
G_0^c(p,q;B_3) \Big[ \tau_{nc}(q;B_3)\mathcal{A}_n(q)\nn
  && +\frac{1}{2} \int_{-1}^1 d\left(\vhat{p}\cdot\vhat{q}\right)
  \tau_{nc}(\pi_3(\vec{p},-\vec{q});B_3)\mathcal{A}_n(\pi_3(\vec{p},-\vec{q}))
\nn
&&+\frac{1}{2} \int_{-1}^1 d\left(\vhat{p}\cdot\vhat{q}\right)
\tau_{nn}(\pi_1(\vec{p},-\vec{q});B_3)\mathcal{A}_c(\pi_1(\vec{p},-\vec{q}))\Big]\,,
\label{eq:Psin-2n}
\end{eqnarray} 
and with the core as a spectator:
\begin{eqnarray}
  \Psi_c(p,q)&=&G_0^n(p,q;B_3) \Big[ \tau_{nn}(q;B_3)\mathcal{A}_c(q)
    \nn
    &&+\int_{-1}^1 d\left(\vhat{p}\cdot\vhat{q}\right)
    \tau_{nc}(\pi_2(\vec{p},-\vec{q});
  B_3)\mathcal{A}_n(\pi_2(\vec{p},-\vec{q}))\big]\,.
\label{eq:Psic-2n}
\end{eqnarray}

With the wave function, we can calculate the one-body matter-density
form factors of an S-wave $2n$ halo.  They are defined as
\begin{eqnarray}
\label{eq:Fy-matter}
F_y(|\bs{k}|)=\int d^3 p \int d^3 q\,
\Psi_y(p,q)~\Psi_y(p,|\vec{q}-\vec{k}|)\,,
\end{eqnarray} 
with $y=n,c$ depending on the Jacobi partitions. For normalized wave
functions $F_y(0)=1$ holds automatically.
The mean-square distance between the valence
neutron and the center of mass of the neutron-core pair,
$\langle r^2 \rangle_{n-nc}$, can be extracted from the form factor $F_n$ via
\begin{equation}
F_n(|\bs{k}|)=1-\frac{1}{6}\bs{k}^2 \langle r^2\rangle_{n-nc}+ \ldots\,,
\end{equation} 
and the mean-square distance between the core and the center of mass of
the two-neutron pair, $\langle r^2 \rangle_{c-nn}$, is determined by
\begin{equation}
F_c(|\bs{k}|)=1-\frac{1}{6}\bs{k}^2\langle r^2\rangle_{c-nn}+ \ldots.
\end{equation}
The geometry of the neutron-neutron-core three-body system
then leads to the following formula for the total matter radius
of a $2n$ halo:
\begin{equation}
\label{eq:rm-3b}
\bra r_m^2 \ket_{2n\rm-halo} 
= \frac{2(A+1)^2}{(A+2)^3} \bra r^2 \ket_{n-nc}
+ \frac{4A}{(A+2)^3} \bra r^2\ket_{c-nn} + \frac{A}{A+2} \bra r_m^2 \ket_{\rm core},
\end{equation}
where the last term is the correction from the finite matter radius of the core.

\subsection{\textit{Applications 3: Efimov states and matter radii}}

\label{sec:universality2nhalos}

In the zero-range limit, long-distance observables in three-body systems
are correlated by few-body universality. One example is the Efimov effect
discussed in the introduction,
which is characterized by discrete scale invariance in the three-body system.
In Eq.~\eqref{eq:H0-3bA}, the running of the three-body coupling, which is a
log-periodic function of the ultraviolet cutoff, is characterized by a limit
cycle with a period $\exp(\pi/s_0)$. As a consequence of the limit cycle, the
three-body S-wave bound states in the unitary limit display a geometric
progression. The ratio of three-body binding energies in two consecutive
states is given by $\exp(2\pi/s_0)$.

Discrete scale invariance has been observed in experiments on ultracold atomic
gases, where the atom-atom scattering length is tuned using a magnetic field
in the vicinity of a Feshbach resonance. Near the unitary limit, the
scattering lengths associated with threshold features in atom-dimer collisions
and three-atom recombination are also correlated through the scaling factor
$\exp(\pi/s_0)$, see~\cite{Braaten:2004rn,Naidon:2016dpf} for reviews. 

\begin{figure}[ht]
\centerline{\includegraphics*[width=0.75\linewidth,angle=0,clip=true]{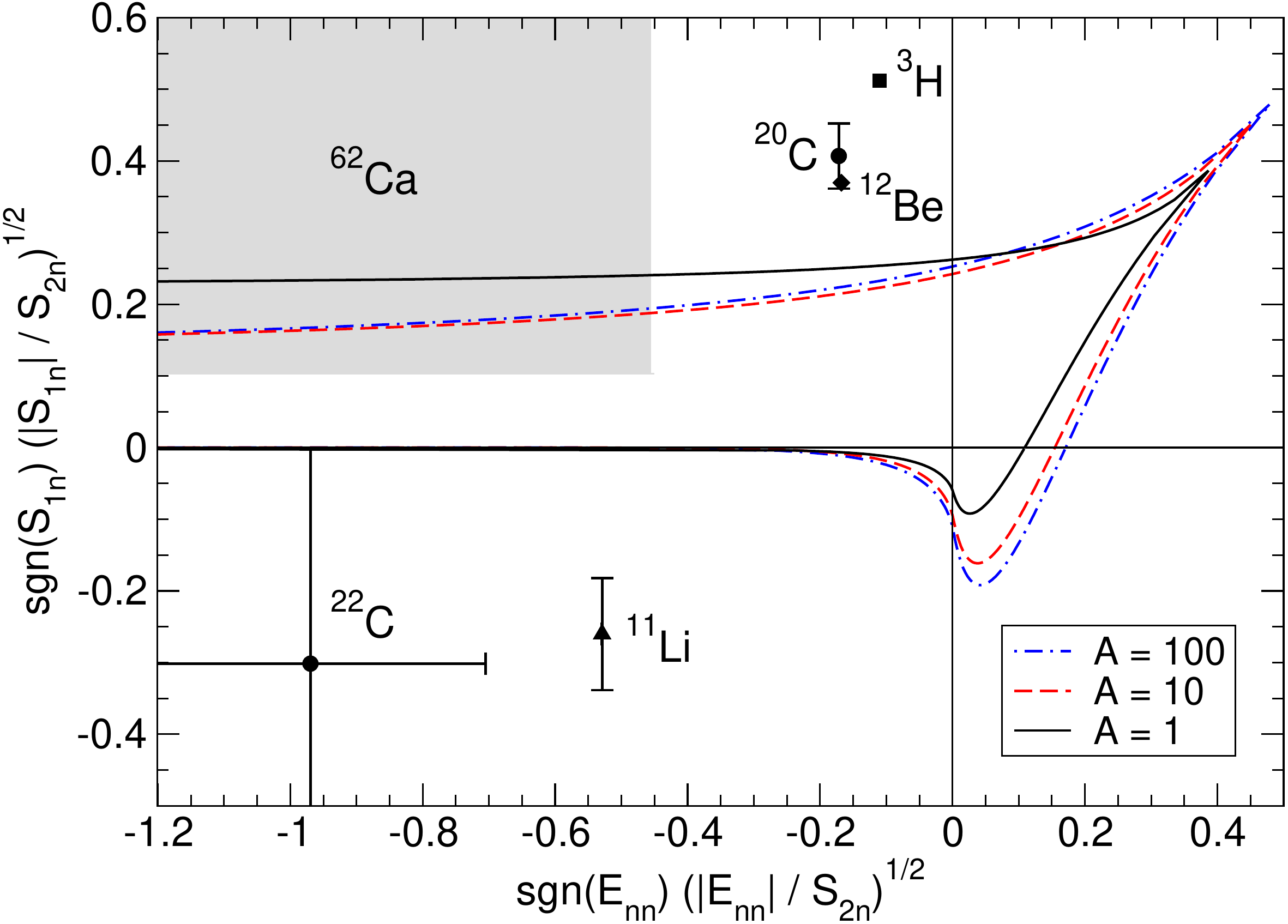}}
\caption{The contour plot in $\textrm{sgn}(E_{nn})\sqrt{|E_{nn}|/S_{2n}}$ versus
  $\textrm{sgn}(S_{1n})\sqrt{|S_{1n}|/S_{2n}}$ for the ground-state $2n$ halos
  with core-neutron mass ratios $A=1,10,100$. The hypothetical bound
  dineutron regime with $E_{nn}>0$ is also included in the theoretical
  calculation to complete the contour.}
\label{pic:halo-efimov}
\end{figure}

In an S-wave $2n$ halo nucleus, $a_{nn}$ and $a_{nc}$ are large but finite,
and the three-body binding energy is characterized by the two-neutron
separation energy, i.e., $S_{2n} = B_3$. In such systems,
the number of possible Efimov-like halo states is determined by the two
ratios $E_{nn}/S_{2n}$ and $S_{1n}/S_{2n}$, where $E_{nn}=-\gamma_{nn}^2/m_n$
is the neutron-neutron virtual energy.
\cite{Amorim:1997mq} suggested to use a universal function
of the ratios $E_{nn}/S_{2n}$ and $S_{1n}/S_{2n}$ to explore possible Efimov
states in halo nuclei, and carried out this study in a zero-range three-body
model (see also~\cite{Yamashita:2007ej}). Following this approach,
\cite{Canham:2008jd} applied EFT to explore the Efimov scenario in S-wave $2n$
halos.
They tuned the three-body coupling $H(\Lambda)$ so that there was an excited
state of the two-neutron halo at threshold, i.e. $B_3^*=\max\{0,E_{nn},S_{1n}\}$.
The $S_{2n}$ of the two-neutron halo is then predicted as all LO two-body and
three-body EFT couplings are fixed. At this value of $S_{2n}$,
LO Halo EFT predicts
the existence of an Efimov excited state at threshold in the halo system.
This, in turn, defines a contour in the $(S_{1n}/S_{2n})$
versus $(E_{nn}/S_{2n})$ plane:
\begin{equation}
g^{(LO)}\left(\frac{E_{nn}}{S_{2n}},\frac{S_{1n}}{S_{2n}};A\right)=1.
\end{equation}
Inside the contour the three-body bound state is deep enough, or the two-body
$nc$ system is close enough to unitarity, for an Efimov state to appear in the
two-neutron halo.  This 
region is depicted in Fig.~\ref{pic:halo-efimov}, and is in good agreement
with an analogous study in a zero-range model~\citep{Frederico:2012xh}.
The curves depend on the core-neutron mass ratio $A$, but
for different values of $A$ quickly
converge to an $A$-independent contour when $A$ increases. 
The specific cases of $^3$H, $^{11}$Li, $^{12}$Be, $^{20,22}$C and $^{62}$Ca are
indicated by mapping their experimental data from
AME2012~\citep{Audi2012,Wang2012} onto this two-dimensional
plane \citep{Hammer:2017tjm}. Note that \cite{Canham:2008jd} originally
came to a different conclusion regarding $^{20}$C since they relied
on older data for $S_{1n}$.

The correlation between bound and excited-state energies is
just one example of the way in which universality imposes relations between
different three-body observables. The point matter radius
$\bra r_m^2\ket_{\rm pt}$ of a ground-state two-neutron halo, defined by
subtracting the core-size contribution from the radius of the halo in
Eq.~\eqref{eq:rm-3b}, is also determined by a universal function of $S_{1n}$
and $S_{2n}$ at LO:
\begin{eqnarray}
\label{eq:point-radii}
\bra r_m^2\ket_{\rm pt}
\equiv \bra r_m^2\ket_{2n\rm-halo} 
-\frac{A}{A+2} \bra r_m^2 \ket_{\rm core}
=\frac{1}{m_n S_{2n}} f^{(LO)}\left(\frac{E_{nn}}{S_{2n}},
\frac{S_{1n}}{S_{2n}};A\right).
\end{eqnarray}
Such correlations have been investigated using EFT for different S-wave
two-neutron halos~\citep{Canham:2008jd,Acharya:2013aea,Hagen:2013jqa}.

$^{22}$C is a Borromean two-neutron halo. The matter radius of $^{22}$C was
determined in the reaction cross section measurement on a proton target to be
$\langle r_m^2\rangle^{1/2}_{2n\rm-halo} =5.4(9)$~fm~\citep{Tanaka:2010zza}. A
more recent interaction cross-section measurement on a carbon target obtained a
more precise result of
$\langle r_m^2\rangle^{1/2}_{2n\rm-halo}=3.44(8)$~fm \citep{Togano:2016wyx},
suggesting a smaller halo configuration in $^{22}$C. 
The $2n$ separation energy of $^{22}$C is not yet directly constrained by
experiment.  In order to obtain an indirect constraint,
\cite{Acharya:2013aea} performed an Halo EFT calculation of the
correlations among $\langle r_m^2\rangle^{1/2}_{2n\rm-halo}$, $S_{1n}$, and $S_{2n}$
of $^{22}$C. Using the value for $\langle r_m^2\rangle^{1/2}_{2n\rm-halo}$
from~\cite{Tanaka:2010zza} known at the time, they predicted an upper
bound on $S_{2n}$ of $0.1$ MeV. \cite{Hammer:2017tjm} updated this
analysis using the more precise value for
$\langle r_m^2\rangle^{1/2}_{2n\rm-halo}$ by \cite{Togano:2016wyx} and
obtained an upper limit of $S_{2n}\le0.4$ MeV in $^{22}$C,
suggesting a more deeply bound system.

$^{62}$Ca, highlighted in Fig.~\ref{pic:halo-efimov} in grey,
is predicted to be a $2n$ halo nucleus,
but to date has not been observed experimentally.
\cite{Hagen:2013jqa} extracted the $n$-$^{60}$Ca scattering parameters from
the $n$--$^{60}$Ca S-wave scattering phase shift, obtained in an
\textit{ab initio} coupled-cluster
calculation based on chiral two- and three-nucleon interactions. The
calculation indicated a large scattering length
$a_{nc}=54(1)$ fm and an effective range $r_{nc}=9.0(2)$ fm for the
$n$-$^{60}$Ca system, where the error is estimated based on the spread of the
coupled-cluster calculations at two small harmonic oscillator
frequencies. This would make ${}^{61}$Ca a very shallow
($S_{1n} \approx 7$ keV) one-neutron halo. 
Using $a_{mc}$ and $r_{nc}$ as input parameters,
they performed a Halo EFT analysis on $^{62}$Ca as a
$n$-$n$-$^{60}$Ca two-neutron halo and searched for possible signatures of
Efimov states. The expansion parameter in this system is estimated
to be $M_{\rm halo}/M_{\rm core}\sim r_{nc}/a_{nc} =1/6$.
They analyzed the LO Halo EFT correlation between the
$n$-$^{61}$Ca scattering length and the two-neutron separation energy $S_{2n}$
of the ground state of $^{62}$Ca. Their conclusion was that for $S_{2n} = 230$
keV an excited state of $^{62}$Ca appears at the $n$-$^{61}$Ca threshold.
Thus for $S_{2n} \geq 230$ keV, ${}^{62}$Ca will have an excited bound state
of Efimovian character. If confirmed in experiment, this would make
$^{62}$C the first $2n$ halo with an Efimov excited state and allow for
a test of the predicted scaling relation.

\subsection{\textit{Range corrections in three-body halos}}
Beyond the leading-order prediction, universal physics in two-neutron halos
is affected by the finite effective range which enters EFT calculations at
next-to-leading order. There are various approaches to include effective
range effects.

The partial resummation technique was developed for studies of range
effects for the triton~\citep{Bedaque:2002yg} and adopted by
\cite{Canham:2009xg} to investigate range corrections in two-neutron
halos. This formalism iterates both the LO and NLO parts of the
two-body scattering amplitudes in the three-body integral equations,
and thus includes some higher-order range corrections above NLO.
These higher-order corrections are small and well-behaved if the
cutoff $\Lambda$ is kept below or close to $M_{\rm core}$.
See \cite{Bedaque:2002yg,Platter:2006ev,Ji:2012nj} for detailed
discussions of this issue.

An alternative, fully perturbative, EFT calculation of range corrections was
recently carried out for two-neutron halos at NLO by
\cite{Vanasse:2016hgn}. This implemented his rigorous method developed
for calculating perturbative range insertions up to N$^3$LO in
the three-nucleon
system~\citep{Vanasse:2013sda,Margaryan:2015rzg,Vanasse:2015fph}. Results for
charge and matter form factors and radii in two neutron halos were obtained
with good accuracy. See \cite{Vanasse:2016hgn,Vanasse:2015fph} for
a detailed discussion of this approach.

\begin{table}
   \centering
   \begin{tabular}{l c c c} \hline
      $\bra r_m^2 \ket_{\rm pt}$ [fm$^2$]
      & $^{11}$Li   
      & $^{14}$Be 
      & $^{22}$C 
      \\  \hline
      EFT$_{\rm LO}$  
      &\quad$5.76\pm2.13$\quad & \quad $1.23\pm0.96$\quad &
      \quad $8.99^{+\infty}_{-5.01}$\quad  
      \\ 
      EFT$_{\rm NLO}$ 
      &\quad$6.16\pm0.84$\quad &\quad$1.40\pm0.85$\quad
      &\quad$9.28^{+\infty}_{-5.17}$\quad
      \\ \hline
      expt
      & \quad$5.34\pm0.15$\quad &\quad $4.24\pm2.42$\quad &\quad
      $21.1\pm9.7$\quad\\
      & & \quad$2.90\pm2.25$\quad &\quad $3.81^{+0.82}_{-0.71}$ \quad
      \\ \hline
   \end{tabular}
   \caption{The point matter radius, Eq.~\eqref{eq:point-radii}, from EFT
     predictions by \cite{Vanasse:2016hgn}
     at LO and NLO.  and from experiments. The experimental
     results for $^{11}$Li and $^{14}$Be are from \citep{Ozawa:2001hb}
     while the experimental results for $^{22}$C are from \citep{Ozawa:2000gx,Tanaka:2010zza} (upper row) and ~\citep{Togano:2016wyx} (lower row).
Compilation taken from \citep{Hammer:2017tjm}.}
\label{tab:rm-2n} 
\end{table}
The matter radii in two-neutron halos were calculated
to LO in Halo EFT by~\cite{Canham:2008jd}. Using partial
resummation, they also predicted the average neutron-core and
neutron-neutron distances in the neutron-neutron-core configuration at NLO
accuracy~\citep{Canham:2009xg}. \cite{Vanasse:2016hgn}
calculated the point matter radii, Eq.~\eqref{eq:point-radii},
at NLO accuracy in the perturbative approach. His results are
fully consistent with the previous calculations
by~\cite{Canham:2008jd,Canham:2009xg}.
In Table~\ref{tab:rm-2n}, we quote Vanasse's results at LO and
NLO and compare with experimental values for $^{11}$Li, $^{14}$Be, and
$^{22}$C. Note that the NLO numbers were obtained by estimating
$r_{nc} \sim1/M_\pi \approx 1.4$ fm. Errors were then
determined as $(r_{nc} \sqrt{2 m_n S_{2n}})^j$, where $j=1$ ($j=2$) for the
LO (NLO) results shown in the first (second) column. 

\section{\textit{Multi-neutron systems}}

Due to the large S-wave neutron-neutron scattering length
$a_{\sing(nn)}= -18.6(5)$~fm  and natural effective
range $r_{\sing(nn)} =2.83(11)$~fm \citep{Chen:2008zzj},
which scale as $1/M_{\rm halo}$ and $1/M_{\rm core}$, respectively,
multi-neutron systems also show universal halo features.
Because $a_{\sing(nn)}$ is negative, the dineutron system is
unbound as was already discussed above in the context of two-body halos.
Here we focus on the universal properties of systems of three and
more neutrons. The expansion parameter $M_{\rm halo}/M_{\rm core}
\approx 1/7$ is rather favorable in this case. Moreover,
three-body interactions are highly suppressed because two of the
neutrons have to be in the same spin state and the Pauli principle
forbids momentum-independent contact interactions.
As a consequence, multi-neutron systems exhibit a higher degree
of universality and only two-body information is required as
input in the first few orders.

The search for multi-neutron resonances and bound states has a
long history~\citep{Kezerashvili:2016ucn}.
The interest in these systems was revived when
experimental evidence for a four-neutron resonance was presented
by~\cite{Kisamori:2016jie}. More recently, \cite{Faestermann:2022meh}
reported indications for a bound tetraneutron and various further experimental
efforts are under way. An overview of the theoretical and experimental
situation can be found in \citep{Marques:2021mqf}. See also \cite{Dietz:2021haj}
for a discussion of the status of three-neutron resonances.

Here, we focus on the universal properties of multi-neutron
systems determined by the ERE parameters $a_{\sing(nn)}$
and $r_{\sing(nn)}$. No further resonant enhancement is assumed.
\cite{Hammer:2021zxb} showed
that such multi-neutron systems display
an approximate non-relativistic \textit{conformal symmetry}
or \textit{Schr\"odinger symmetry} \citep{Hagen:1972pd,Mehen:1999nd}
if their center-of-mass energy $E_{cm}$ is in the range
\begin{equation}
  \label{eq:neutrons}
  0.1\mbox{ \textrm MeV}\approx \frac{1}{m_n (a_{\sing(nn)})^2}
  \ll E_{cm} \ll
  \frac{1}{m_n (r_{\sing(nn)})^2}\approx 5\mbox{ \textrm MeV}\,,
\end{equation}
and elucidated the consequences of this symmetry for
multineutron systems. For neutrons in this energy range,
one can safely set $1/a_{\sing(nn)}=0$ and $r_{\sing(nn)}=0$ at LO
in the EFT description, such that no dimensionful interaction parameters
are left and the theory is scale invariant.
The multi-neutron systems can then be described by a field $\mathcal{U}$
in a non-relativistic \textit{conformal field theory}
(CFT) \citep{Nishida:2007pj}.
In addition to the usual Galilei transformations, the CFT is also
invariant under scale transformations
\begin{equation}
  \bs{x} \to e^\lambda \bs{x},\qquad t \to  e^{2\lambda}t\,,
\end{equation}
and special conformal
transformations
\begin{equation}
\bs{x} \to \frac{\bs{x}}{1+\xi t}\,,\qquad t \to
\frac{t}{1+\xi t}\,,
\end{equation}
of the space-time variables $\bs{x}$ and $t$, where $\lambda$ and $\xi$
are real parameters.

The full Green's function of a field $\mathcal{U}$ in CFT is strongly constrained
by the conformal symmetry. Up to an overall constant, it depends only on
the mass of the field $M_\mathcal{U}$ and the scaling dimension
$\Delta$, which determines the behavior of the field under
scale transformations. \cite{Hammer:2021zxb} used this
property to derive
a general expression for the energy spectra of multi-neutron
systems created in high-energy nuclear reactions in terms of $\Delta$.
Consider a nuclear reaction of two nuclei $A_1$ and $A_2$, which
produces a few final-state neutrons and a recoil particle $B$, 
\begin{equation}
  A_1 + A_2 \to B + \underbrace{n + n + \cdots}_{N~\text{neutrons}}\,.
\end{equation}
The number of final-state neutrons $N$ can be $2, 3, 4,\ldots$.
The energy scale of the primary nuclear reaction
that produces the neutrons is assumed to be large
compared to $E_{cm}$,
such that the corresponding matrix element factorizes into
two parts: one describing the primary production process
which creates the multi-neutron system in a point
and one corresponding to the final-state interaction of the neutrons.
In the case of two neutrons this provides the basis for the Watson-Migdal
approach to final-state interactions~\citep{Watson:1952ji,Migdal:1955}.

Consider first an experiment where only the energy of
the recoil particle $B$ is measured. The inclusive
differential cross section as function of the recoil energy $E$
of particle $B$, $d\sigma/dE$, is continuous and vanishes at some maximum
recoil energy $E_0$.
Near the end point, where the neutrons satisfy Eq.~\eqref{eq:neutrons},
it has the general form 
\begin{equation}
  \frac{d\sigma}{dE} \sim (E_0-E)^{\Delta-\frac{5}{2}}\,.
\end{equation}
If the low-energy spectrum $P(E_{cm})$ of the multineutron system in its
center-of-mass itself is measured, it takes the form
\begin{equation}
  \label{eq:P-neutron}
  P(E_{cm})\sim (E_{cm})^{\Delta-\frac{5}{2}}\,.
\end{equation}
Thus the energy spectra for $E_{cm}$ in the kinematical range given by
Eq.~\eqref{eq:neutrons} follow a universal power law that is
determined by the scaling dimension of the $N$-neutron system.
For very small energies $\sqrt{m_n E_{cm}} \ll 1/a_{\sing(nn)}$,
the neutrons show free particle behavior. The transition to the conformal
window can be calculated in pionless EFT.

The scaling dimension $\Delta$ for a free neutron is $3/2$, for
$N$ free neutrons it is $3N/2$. For a system of interacting
neutrons $\Delta$  can be obtained from  a field theory calculation
(cf.~\cite{Braaten:2021iot})
or from the energy of the corresponding
few-neutron system in the unitary limit placed
in a harmonic potential with unit oscillator
frequency \citep{Nishida:2007pj}. This so-called
\textit{operator state correspondence}
leads to an nontrivial connection between the few-body physics
of spin-1/2 fermions at unitarity and the physics of nuclear reactions.
Namely, the spectrum of spin-1/2 fermions at unitarity in a harmonic
trap determines the behavior of the processes involving
emission of neutrons in a certain kinematic regime.
The lowest scaling dimensions for some $N$-neutron systems are
collected in Table~\ref{tab:scaling}.

\begin{table}
  \centering
\begin{tabular}{c c c c}
\hline
$N$\quad &\quad $S$\quad & \quad$L$\quad &\quad $\Delta$\\
\hline
2\quad &\quad 0\quad &\quad 0\quad &\quad 2\\
3 \quad&\quad 1/2 \quad&\quad 0\quad &\quad 4.66622\\
3 \quad&\quad 1/2 \quad&\quad 1\quad &\quad 4.27272\\
4 \quad&\quad 0\quad &\quad 0\quad &\quad 5.07(1)\\
5 \quad&\quad 1/2 \quad&\quad 1 \quad&\quad 7.6(1) \\
6 \quad&\quad 0 \quad&\quad 0 \quad&\quad 8.67(3) \\
\hline
\end{tabular}
  \caption{Scaling dimensions $\Delta$ for $N$-neutron systems
    with spin $S$ and orbital angular momentum $L$ in the
    unitary limit. Numbers taken from \citep{Nishida:2007pj}.}
\label{tab:scaling} 
\end{table}

\cite{Hammer:2021zxb} demonstrated that the
two- and three-neutron spectra of state-of-the-art
calculations for the reactions
$^6$He$(p,p\alpha)2n$~\citep{Gobel:2021pvw},
$^3$H$(\pi^-,\gamma)3n$~\citep{Golak:2018jje},
and $^3$H$(\mu^-,\nu_\mu)3n$~\citep{Golak:2016zcw}
exhibit the universal power law behavior. Moreover, the power law
is consistent with the experimental photon spectrum near the
kinematical end point for radiative capture of stopped pions on
tritium measured by~\cite{Miller:1980pn}. A stringent test of the
prediction, Eq.~\eqref{eq:P-neutron}, may become possible in
current experiments
searching for tetraneutron resonances which will produce precise
four-neutron spectra at low energies.

\section{\textit{Further reading}}

In this chapter, we have provided an introduction to the
description of two and three-body halo systems using Halo EFT.
Due to space constraints, the focus has been on systems
dominated by S-wave neutron-core interactions and their
universal structural properties.

Halo EFT and universality has a much wider range of applications,
including systems with higher partial wave interactions
and/or strangeness, proton halos and systems with Coulomb repulsion,
as well as electroweak currents and reactions.
For a detailed discussion of these issues, we refer
the reader to the reviews by \cite{Rupak:2016mmz,Hammer:2017tjm,Hammer:2019poc}.
An application of Halo EFT to describe nuclear reactions was recently developed
by~\cite{Capel:2018kss,Moschini:2019tyx,Capel:2020obz}.

For the description of the properties and reactions of
halo nuclei using similar and complementary
methods, we refer the reader to the excellent reviews on this topic
\cite{Zhukov:1993aw,Nielsen2001,Jensen:2004zz,Ogata:2009cx,Baye:2010ht,Pfutzner:2011ju,Frederico:2012xh,Canto:2015esm,Moro:2018mdk}.

We thank Chen Ji and Daniel R. Phillips for a fruitful collaboration that
led to this chapter and Chen Ji for comments on the manuscript.
This work was supported in part by the
Deutsche Forschungsgemeinschaft (DFG, German Research Foundation) –
Projektnummer 279384907 – SFB
1245 and by the German Federal Ministry of Education
and Research (BMBF) (Grant No. 05P21RDFNB).

\bibliographystyle{apsrmp4-2}
\bibliography{HTrefs}
\end{document}